%% file: main.tex
\documentclass[sigconf,balance=false]{acmart}

\usepackage{popets}
\setcopyright{popets}
\copyrightyear{YYYY}

\acmYear{YYYY}
\acmVolume{YYYY}
\acmNumber{X}
\acmDOI{XXXXXXX.XXXXXXX}
\acmISBN{}
\acmConference{Proceedings on Privacy Enhancing Technologies}
\usepackage{xspace}
\usepackage{naveed}
\usepackage{hamza}

\begin{document}


\input{title}
\input{abstract}


\maketitle

\input{introduction}
\input{related_works_2}

\input{preliminaries}
\input{security_model}

\input{protocols}
\input{security_analysis}
\input{privacy_preserving_ml}
\input{evaluation}
\input{conclusion}

\bibliographystyle{ACM-Reference-Format}
\bibliography{refs}
\input{appendix}

\end{document}

%% file: title.tex
\title[Hawk: Accurate and Fast PPML Using Secure Lookup Table Computation]{Hawk: Accurate and Fast Privacy-Preserving Machine Learning Using Secure Lookup Table Computation}

\author{Hamza Saleem}
\affiliation{%
  \institution{University of Southern California}
  \city{} 
  \country{}
  }
\email{hsaleem@usc.edu}

\author{Amir Ziashahabi}
\affiliation{%
  \institution{University of Southern California}
  \city{}
  \country{}}
\email{ziashaha@usc.edu}

\author{Muhammad Naveed}
\affiliation{%
  \institution{University of Southern California}
  \city{}
  \country{}}
\email{mnaveed@usc.edu}

\author{Salman Avestimehr}
\affiliation{%
  \institution{University of Southern California}
  \city{}
  \country{}}
\email{avestime@usc.edu}

\renewcommand{\shortauthors}{Saleem et al.}

%% file: abstract.tex
\begin{abstract}
%-------------------------------------------------------------------------------
Training machine learning models on data from multiple entities without direct data sharing can unlock applications otherwise hindered by business, legal, or ethical constraints. In this work, we design and implement new privacy-preserving machine learning protocols for logistic regression and neural network models. We adopt a two-server model where data owners secret-share their data between two servers that train and evaluate the model on the joint data. A significant source of inefficiency and inaccuracy in existing methods arises from using Yao’s garbled circuits to compute non-linear activation functions. We propose new methods for computing non-linear functions based on secret-shared lookup tables, offering both computational efficiency and improved accuracy.

Beyond introducing leakage-free techniques, we initiate the exploration of relaxed security measures for privacy-preserving machine learning. Instead of claiming that the servers gain no knowledge during the computation, we contend that while some information is revealed about access patterns to lookup tables, it maintains $\epsilon$-$d_{\mathcal{X}}$-privacy. Leveraging this relaxation significantly reduces the computational resources needed for training. We present new cryptographic protocols tailored to this relaxed security paradigm and define and analyze the leakage.
Our evaluations show that our logistic regression protocol is up to $9\times$ faster, and the neural network training is up to $688\times$ faster than SecureML~\cite{secureml}. Notably, our neural network achieves an accuracy of $96.6\%$ on MNIST in 15 epochs, outperforming prior benchmarks~\cite{secureml,securenn} that capped at $93.4\%$ using the same architecture.

\end{abstract}

%% file: introduction.tex
\section{Introduction}

Machine learning (ML) has become an indispensable tool in various domains, from advertising and healthcare to finance and retail, enabling the training of predictive models. The accuracy of these models is significantly enhanced when trained on vast datasets aggregated from diverse sources. 
%For instance, a consortium of hospitals might wish to merge their patients' sensitive data to develop a diagnostic model with superior disease prediction capabilities. 
However, numerous challenges, such as privacy concerns, stringent regulations, and competitive business landscapes, often hinder collaborative data-sharing endeavors.

Privacy-Preserving Machine Learning (PPML) using secure computation allows multiple entities to train models on their aggregated data without revealing their individual datasets. In this paradigm, data owners distribute their private data among several non-colluding servers using secret-sharing techniques. These servers then perform training on the combined data without gaining insights into the individual datasets.

Past works on PPML have focused on training logistic regression models~\cite{logistic_regression_13_paper, secureml} and neural networks~\cite{secureml, securenn, ABY_paper,mohassel2018aby3}. A significant bottleneck in existing work is using expensive protocols, such as Yao's garbled circuits, for computing non-linear activation functions. Furthermore, prior works~\cite{secureml, securenn} resort to non-standard or approximate activation functions, resulting in accuracy loss.

In this work, we present two novel protocols $\textsc{Hawk}_{\text{Single}}$ and $\textsc{Hawk}_{\text{Multi}}$, designed for accurate and efficient computation of activation functions and their derivatives. These protocols leverage secret shared precomputed lookup tables. We begin by introducing the $\textsc{Hawk}_{\text{Single}}$ protocol, which offers plaintext accuracy and robust security in a semi-honest adversary model. Its primary limitation, however, is its reliance on consuming one lookup table for each function computation, necessitating a vast number of such tables. 
To mitigate this limitation, we subsequently introduce $\textsc{Hawk}_{\text{Multi}}$, enabling lookup table reusability. The $\textsc{Hawk}_{\text{Multi}}$ protocol poses several security and efficiency challenges, which we further address to achieve a fast and secure protocol. Although methods such as Oblivious RAM (ORAM) \cite{goldreichSoftwareProtectionSimulation1996} can be employed to eliminate information leakage in our context, we opt for a slightly relaxed security model to prioritize performance improvements. Our $\textsc{Hawk}_{\text{Multi}}$ protocol permits table reuse but at the cost of revealing limited information about the access patterns. We rigorously analyze the leakage of the $\textsc{Hawk}_{\text{Multi}}$ protocol and prove that it preserves $\epsilon$-$d_{\mathcal{X}}$-privacy for the leaked access patterns.

% In this work, we introduce two novel protocols $\textsc{Hawk}_{\text{Single}}$ and $\textsc{Hawk}_{\text{Multi}}$ to accurately and efficiently compute activation functions and their derivatives. These protocols leverage secret shared precomputed lookup tables. While the $\textsc{Hawk}_{\text{Single}}$ protocol offers robust security in a semi-honest adversary model, it consumes one lookup table for each function computation, necessitating a vast number of such tables.
% $\textsc{Hawk}_{\text{Multi}}$ protocol permits table reuse but at the cost of revealing minimal information about the access patterns to these tables. Although fully oblivious methods, such as Oblivious RAM (ORAM), can be employed to eliminate
% Lookup table reuse in $\textsc{Hawk}_{\text{Multi}}$ protocol poses several security and efficiency challenges, which we address to achieve a fast and secure protocol. We rigorously analyze the leakage of $\textsc{Hawk}_{\text{Multi}}$ protocol and prove that it preserves $\epsilon$-$d_{\mathcal{X}}$-privacy for the leaked access patterns. While techniques like Oblivious RAM (ORAM) could be employed to eliminate such leakage, they are often deemed inefficient for our use case.

Building on $\textsc{Hawk}_{\text{Single}}$ and $\textsc{Hawk}_{\text{Multi}}$ we develop efficient PPML protocols tailored for logistic regression and neural network training in a two-server secure computation environment. Our evaluations, conducted on Amazon EC2, showcase the performance gains of our system.

\subsection{Our Contributions}

%The following are our main contributions:
\sloppy \noindent \textbf{Computing Standard Activation Functions.} We propose a novel method to accurately compute activation functions in a secure computation setting and achieve accuracy similar to plaintext training. Past methods either use non-standard activations or approximate them. As shown in Figure~\ref{fig:convergence_comparison_NN}, using non-standard activation functions may cause accuracy to plummet to random guess levels, potentially affecting convergence. Hence, using standard activation functions is crucial to harness machine learning advancements rather than relying on makeshift functions for secure computation.

\noindent \textbf{Generic Method for Precise Univariate Function Computation.} We introduce a universal method for securely computing any univariate function.  It can be directly used to compute any univariate activation function, which constitutes most of the activation functions used in all but the last layer of a neural network. Our method computes activations much faster than the state-of-the-art. This method creates secret shared lookup tables for all possible inputs; we use recent advances in low-precision machine learning to confine the size of the lookup tables.

\noindent \textbf{Precise Computation of Multivariate Functions.}
We demonstrate the application of our method to compute multivariate activations, notably $\softmax$. We highlight that this is the first method to compute $\softmax$ in a secure computation setting accurately.

\noindent \textbf{Beyond Activation Functions.}
While we demonstrate the applicability of our protocols for activation function computation, their potential extends far beyond these specific functions. The underlying principle of precomputed lookup tables can be applied to a wide range of MPC applications where the input domain can be bounded. For example, in private set intersection~\cite{psi_paper}, our protocol can enable efficient comparison of elements in private sets. Similarly, secure auctions~\cite{secure_auctions} can benefit from lookup tables to ensure privacy while evaluating bids. Furthermore, our protocols can be used to securely compute histograms~\cite{mazloomSecureComputationDifferentially2018b}, enabling private data analysis. These applications highlight the versatility of our protocols for MPC across various domains.

\sloppy \noindent \textbf{An Alternate Representation for $\mathsf{\textbf{DReLU}}$ Activation.} We present an alternate representation of the $\drelu$ activation function for fixed-point numbers, which allows us to reduce the entries in the lookup table for $\drelu$ by a factor of $2^{11}$.

\noindent \textbf{Relaxed Security Setting.} 
In addition to introducing leakage-free methods using $\hawksingle$, we initiate the study of a new security model for PPML. This relaxed model allows some information leakage about access patterns, but the leakage is proven to preserve $d_{\mathcal{X}}$-privacy. We leverage this model to develop $\hawkmulti$ protocol to significantly reduce computational resources while training.

\noindent \textbf{Experimental results.} We implement our framework and perform various experiments for training and inference of logistic regression and neural networks using six different datasets in both LAN and WAN settings. Section~\ref{sec:evaluation} shows that our protocols are significantly faster than state-of-the-art in a two-party setting and achieve accuracy similar to plaintext training. 

%% file: related_works_2.tex
\section{Related Work}
\label{sec:related-work}
%Our protocol leverages lookup tables to address prevalent inefficiencies of PPML in secure multi-party computation. 
This section overviews prior studies across fields such as access pattern privacy, PPML, and secure MPC pertinent to our work.

\noindent \textbf{Access Pattern Privacy} literature focuses on achieving full or partial obliviousness during data retrieval. ORAM
\cite{goldreichSoftwareProtectionSimulation1996} algorithms provide complete obliviousness by ensuring that the physical access patterns for any two equal-length logical patterns are indistinguishable. While ORAM addresses access pattern leakage and has been extensively studied in both client-server \cite{goldreichSoftwareProtectionSimulation1996, shi2011oblivious, stefanov2018path, asharov2020optorama, pinkas2010oblivious} and multi-party \cite{doernerScalingORAMSecure2017, wang2014scoram, zahurRevisitingSquareRootORAM2016, pinkas2010oblivious, gordon2012secure} settings, it carries an inevitable overhead of $\Omega(log(n))$ \cite{goldreichSoftwareProtectionSimulation1996, boyle2016there, larsen2018yes}. 

A natural approach for circumventing this overhead is relaxing full obliviousness for a weaker privacy notion, such as differential privacy. Differential privacy introduces a neighborhood concept and ensures closeness between neighboring access sequences rather than all sequences. Despite the relaxation, \cite{persiano2023lower} show that the $\Omega(log(n))$ lower bound also applies to DP-ORAM \cite{wagh2018differentially} structures. Nonetheless, various studies have employed differentially private access patterns to offer performance enhancements for specific algorithms, such as searchable symmetric encryption (SSE)~\cite{shang2021obfuscated}, 
sorting and merging of lists \cite{chanFoundationsDifferentiallyOblivious2022a}, histograms \cite{allen2019algorithmic}, and graph-parallel computation \cite{mazloomSecureComputationDifferentially2018b}. The performance enhancement for these algorithms stems from the fact that they generally rely on a full memory scan. Therefore, the problem of access pattern privacy is reduced to hiding the number of accesses for each element. For lookup tables, where only a limited number of elements are accessed, these algorithms fail to deliver an efficient solution.

\noindent \textbf{PPML Using Secure MPC}.
Previous MPC studies have explored decision trees~\cite{decision_trees}, K-means clustering~\cite{k_means2, k_means1}, SVM~\cite{svm1, svm2}, and regression techniques~\cite{linear_reg1, linear_reg2, linear_reg3, logistic_reg}. However, these works incur high efficiency overheads and lack implementation. SecureML~\cite{secureml} provides protocols for secure training and inference of linear regression, logistic regression, and neural networks in a two-server setting. They present alternate functions to approximate the non-linear activations ($\sigmoid$, $\relu$, and $\softmax$), which are evaluated using garbled circuits. However, garbled circuits have a high communication overhead, and approximating activation functions reduces the model's accuracy.
QUOTIENT~\cite{quotient_paper} presents protocols for neural network training in a two-server setting, utilizing quantization for various network components. However, using garbled circuits and oblivious transfer produces high overheads.

SecureNN~\cite{securenn} and ABY$^3$~\cite{mohassel2018aby3} present privacy-preserving neural network training protocols in a three-server honest majority setting. Like SecureML, they approximate $\softmax$, which reduces the model accuracy, as discussed in Section~\ref{subsec:neural_network_training}. Falcon~\cite{falcon_paper} combines techniques from SecureNN~\cite{securenn} and ABY$^3$~\cite{mohassel2018aby3} to improve the training performance. Our two-server protocol, as shown in Section~\ref{sec:evaluation}, outperforms these in efficiency. Another pertinent work in the three-server setting is Pika ~\cite{Wagh_2022}, which adopts a lookup table-centric approach to computing non-linear functions. Pika relies on Private Information Retrieval (PIR) and Function Secret Sharing (FSS)~\cite{boyle2015function} to hide access patterns. Despite providing full obliviousness, their protocol has an overhead of $O(n)$ stemming from FFS key generation, FFS evaluation, and dot product operation with all lookup table elements. In fact, $\Omega(n)$ overhead is intrinsic to PIR~\cite{Beimel2004ReducingTS}.
Pika only provides performance evaluations for generic functions and inference tasks. We compare our protocols with Pika for sigmoid function computation in Appendix~\ref{app:comparison}.

A parallel line of work focuses on secure neural network inference. A party or a group of parties evaluates the model on a client's private data, keeping the model private. Gilad-Barach et al.\cite{cryptonets} employ HE for secure inference, albeit with accuracy compromises. 
%They approximate the $\relu$ activation function using a quadratic function, which results in accuracy loss.
MiniONN~\cite{mini_onn} optimizes the secure inference protocols of SecureML~\cite{secureml} by reducing the offline cost of matrix multiplication.
%while increasing the online cost. 
Using a third-party dealer, Chameleon~\cite{chameleon} removes the oblivious transfer protocols required for multiplication. Gazelle~\cite{gazelle} optimizes linear operations using specialized packing schemes for HE. Baccarini et al.~\cite{Baccarini2023MultiPartyRS} present ring-based protocols for MPC operations applied to NN inference. SIRNN~\cite{sirnn} presents a lookup table based protocol that uses digit decomposition with oblivious transfers to compute non-linear functions efficiently. We compare our protocols with SIRNN for sigmoid function computation in Appendix~\ref{app:comparison}.

\noindent \textbf{Other PPML Approaches}.
Shokri and Shmatikov~\cite{priv_deep_learning} explore privacy-preserving neural network training with horizontally partitioned data. The servers train separately and share only parameter updates, using differential privacy~\cite{diff_priv_book} to minimize leakage. Wu et al.\cite{logistic_regression_13_paper} and Kim et al.\cite{kim_he_logistic} employ homomorphic encryption for logistic regression, approximating the logistic function with polynomials. The complexity is exponential in the polynomial degree, and the approximation reduces the model accuracy. Crawford et al.~\cite{FHE_LR_Lookup} use fully homomorphic encryption to train logistic regression models using encrypted lookup tables. However, the approach incurs high overhead, and using low-precision numbers reduces the model's accuracy. In contrast, our protocols are orders of magnitude more efficient and achieve accuracy similar to plaintext training.

We primarily compare our techniques with SecureML~\cite{secureml} as this is the only work we found for logistic regression and neural network training in a two-server client-aided setting. While recent works exist~\cite{quotient_paper, securenn, mohassel2018aby3, falcon_paper}, their settings differ, making direct comparisons challenging. We compare our online cost with theirs as they lack an offline phase. We further remark that the idea of employing lookup tables for function evaluation is not novel with past works~\cite{tinytable_paper, lookup_table_paper_Dessouky, lookup_table_paper_Keller, boyle2015function, sirnn} suggesting circuit and FSS-based protocols. However, these works necessitate representing functions as either boolean or arithmetic circuits, with each gate evaluated using a lookup table. It is non-trivial to directly represent functions like exponential, logarithm, etc, used in ML activation, as circuits. This complexity typically leads to reliance on circuit-friendly approximations, such as the Taylor series, which compromises accuracy and introduces inefficiencies (see Figure~\ref{fig:convergence_comparison_NN}). Our lookup table based protocols are fundamentally different as they enable direct computation of ML activations without resorting to circuit approximations. Furthermore, our main contribution is the $\textsc{Hawk}_{\text{Multi}}$ protocol that allows for the reuse of a lookup table by leveraging metric differentially private access patterns, missing in prior research.

%% file: preliminaries.tex
\section{Preliminaries}

\subsection{Machine Learning}

\noindent \textbf{Logistic Regression} is a binary classification algorithm commonly, often used for tasks like medical diagnosis. Given $n$ training examples $\boldsymbol{x_1, x_2, ..., x_n}\in \mathbb{R}^D$ with binary labels $y_1, y_2, ..., y_n$, the goal is to learn a coefficient vector $\boldsymbol{w}\in \mathbb{R}^D$ which minimizes the distance between $g(\boldsymbol{x_i}) = f(\boldsymbol{x_i.w})$ and the true label $y_i$. The \textit{activation function} $f$ is used to bound the dot product of $\boldsymbol{x_i}$ and $\boldsymbol{w}$ between 0 and 1 for binary classification. The $\sigmoid$ function is typically chosen for $f$, defined as: $f(z) = \frac{1}{1+e^{-z}}$. To learn $\boldsymbol{w}$, we define a cost function $C(\boldsymbol{w})$ and find $\boldsymbol{w}$ using the optimization $\argmin_{\boldsymbol{w}}C(\boldsymbol{w})$. We use the cross-entropy cost function defined as: $C_i(\boldsymbol{w}) = -y \log f(\boldsymbol{x_i.w}) - (1-y)\log(1-f(\boldsymbol{x_i.w}))$, with the overall cost being the average across all examples: $C(\boldsymbol{w}) = \frac{1}{n} \sum_{i=1}^{n} C_i(\boldsymbol{w})$.    

\noindent \textbf{Neural Networks} extend logistic regression to handle more complex tasks. A neural network consists of $l$ layers where each layer $i$ contains $d_i$ neurons. The output of each neuron is computed by applying a non-linear activation to the dot product of its coefficient vector $\boldsymbol{w}$ and the input $\boldsymbol{x}$. Activation functions, like the ReLU: $f(z) = max(0,z)$, are commonly used. The first layer receives inputs from the dataset, with each subsequent layer performing calculations and forwarding the output to the next layer using ‘forward propagation’. For classification, the probability of the input example belonging to each class is computed by applying the $\softmax$ function: $f(z_i) = \frac{e^{z_i}}{\sum_{i=1}^{d_l}e^{z_i}}$ to each neuron in the output layer, where $d_l$ denotes the total neurons in the output layer. The $\softmax$ output is a probability value between 0 and 1 for each class, summing up to 1. To train the network, we apply the Stochastic Gradient Descent (SGD) algorithm and update the coefficients of all neurons recursively, starting from the last layer and moving towards the first layer in a `backward propagation' step. For further details on related ML concepts like SGD and batching, refer to Appendix~\ref{app:machine_learning}.

\subsection{Secure Computation}
\label{sec:prelim-secure-computation}

Secure computation allows parties to compute a joint function of their data while keeping their data private. 

\noindent \textbf{Preprocessing Model of Secure Computation} divides a protocol into two phases: a data-independent offline phase used to create correlated randomness and an online phase. This model allows most of the heavy computation to be delegated to the offline preprocessing phase, rendering the online phase more computationally efficient.

\noindent \textbf{Additive Secret-Sharing}. In a secure two-party computation (2PC) setting, secret values are additively secret-shared between the parties $P_0$ and $P_1$. To share a secret $x \in \mathbb{Z}_{n}$, a random $r \in \mathbb{Z}_{n}$ is generated. The share for the first party is computed as $\share{x}{0} = r$, while the second party's share is $\share{x}{1} = x - r \pmod{n}$. To reconstruct $x$, one party sends its share to the other party, which then computes $x = \share{x}{0} + \share{x}{1} \pmod{n}$, denoted as $Rec(x_0, x_1)$. We use $\share{x}{i}$ to denote $P_i$'s share of $x$. 

\noindent \textbf{Secure Addition.} The parties can non-interactively add two secret-shared values $x$ and $y$ where each $P_i$ computes $\share{z}{i} = \share{x}{i} + \share{y}{i} \pmod{n}$ as their share of the sum. 

\noindent \textbf{Secure Multiplication.} To multiply two secret-shared values $x$ and $y$, the parties use Beaver's multiplication triples~\cite{beaver1991efficient}. Initially, the parties possess shares of $a$, $b$, and $c$ where $a$ and $b$ are random values in $\mathbb{Z}_{n}$ and $c = ab \pmod{n}$. Each party $P_i$ computes $\share{d}{i} = \share{x}{i} - \share{a}{i} \pmod{n}$ and $\share{e}{i} = \share{y}{i} - \share{b}{i} \pmod{n}$ and compute $d = Rec(d_0, d_1)$ and  $e = Rec(e_0, e_1)$ to reconstruct $d$ and $e$. Each party $P_i$ then computes $\share{z}{i} = \share{x}{i} e + \share{y}{i} d + \share{c}{i} - i d e$ as their share of the product. 

\subsection{\boldmath Metric Differential Privacy ($d_{\mathcal{X}}$-privacy)}
\label{sec:metric_differential_privacy}
\begin{definition}
A randomized algorithm $\mathcal{L}$ is $\epsilon$-differentially private if for all datasets $D_1$ and $D_2$ differing in a single value and for all $S \subset Range(\mathcal{L})$:
{\footnotesize
$Pr[\mathcal{L}(D_1) \in S] \leq e^{\epsilon} Pr[\mathcal{L}(D_2) \in S]$}, where the probability space is over the coin flips of the mechanism $\mathcal{L}$.
\end{definition}

While the above definition captures differential privacy in a centralized setting, there are scenarios where data privacy at the individual level is paramount. Local Differential Privacy (LDP) obfuscates individual data records when clients distrust the central curator, altering data before statistical analysis.

\begin{definition}
A randomized algorithm $\mathcal{L}$ provides $\epsilon$-LDP if for each pair of inputs $x$ and $x'$ and for all $S \subset Range(\mathcal{L})$:
{\footnotesize
$Pr[\mathcal{L}(x) \in S] \leq e^{\epsilon} Pr[\mathcal{L}(x') \in S]$}.
\end{definition}

In this work, we adopt a natural generalization of the conventional differential privacy~\cite{diff_priv_book} presented by Chatzikokolakis et al.~\cite{d_privacy_paper}. This extension is particularly relevant when the domain $\mathcal{X}$ is viewed as a metric space. Termed as metric differential privacy or $d_{\mathcal{X}}$-privacy, this approach is adept at modeling situations where indistinguishability is best expressed through a metric between secret inputs. 
Geo-Indistinguishability is presented as a canonical example for $d_{\mathcal{X}}$-privacy~\cite{d_privacy_paper, geo_location_priv_paper}, where a user wishes to share their location with a service to receive nearby restaurant recommendations. Instead of disclosing the precise coordinates, the user can employ $d_{\mathcal{X}}$-privacy to share an \textit{approximated location} to receive relevant suggestions without compromising their exact whereabouts. 

The input domain $\mathcal{X}$ for an activation function can be viewed as a metric space. For instance, if for some $x,x_1,x_2 \in \mathcal{X}$, $|x-x_1|<|x-x_2|$, then usually, $f(x_1)$ provides a better approximation of $f(x)$ than $f(x_2)$. 
%Hence, local $d_{\mathcal{X}}$-privacy is a more appropriate primitive for our use case than the conventional local DP. 
Mechanisms from standard or local DP can be adapted to suit $d_{\mathcal{X}}$-privacy by choosing a fitting metric for the confidential data. We leverage local $d_{\mathcal{X}}$-privacy to obfuscate each individual table lookup in our $\hawkmulti$ protocol, ensuring enhanced privacy from computing servers.

\begin{definition}
\label{def:local_dx_privacy}
A randomized algorithm $\mathcal{L}$ provides local $\epsilon$-$d_{\mathcal{X}}$-privacy if for each pair of inputs $x$ and $x'$ and for all $S \subset Range(\mathcal{L})$: {\footnotesize \noindent $Pr[\mathcal{L}(x) \in S] \leq e^{\epsilon.d_{\mathcal{X}}(x,x')} Pr[\mathcal{L}(x') \in S]$}, where $d_{\mathcal{X}}(.,.)$ is a distance metric.
\end{definition}

\noindent 
%The metric $d_{\mathcal{X}}(.,.)$ should satisfy the axioms:  $d_{\mathcal{X}}(x_a,x_a) = 0$, $d_{\mathcal{X}}(x_a,x_b) \geq 0$, $d_{\mathcal{X}}(x_a,x_b) = d_{\mathcal{X}}(x_b,x_a)$, $d_{\mathcal{X}}(x_a,x_b) + d_{\mathcal{X}}(x_b,x_c) \geq d_{\mathcal{X}}(x_a,x_c)$, with $x_a, x_b, x_c \in \mathcal{X}$. 
We define the distance metric as the absolute difference between two points, i.e., $d_{\mathcal{X}}(x,x') = |x - x'|$. We choose $d_{\mathcal{X}}$-privacy as the noising mechanism to obfuscate table lookups while ensuring the noised value remains close to the original. This proximity is crucial for maintaining model accuracy. By definition, $d_{\mathcal{X}}$-privacy offers a stronger privacy guarantee for inputs that are close to each other due to similar output distributions after adding noise. We can achieve a stronger guarantee for distant input values by adding large noise but at the cost of utility. Our protocol aims to balance privacy, accuracy, and efficiency. Opting for stronger privacy guarantees and higher accuracy would require more lookup tables, each with a smaller privacy budget but at a higher offline phase and storage costs. Conversely, optimizing for privacy and efficiency would necessitate adding more noise to each lookup, potentially compromising accuracy. Lastly, for leakage analysis, it is common in related studies~\cite{mazloomSecureComputationDifferentially2018b} to focus on access pattern leakage. Analyzing potential leaks for the underlying input data tends to be quite complex.

%\subsection{Geometric Mechanism}
%\label{sec:geometric_mechanism}
Differential privacy employs various mechanisms to introduce controlled randomness to query results. In general, mechanisms designed for differential privacy can be extended to $d_{\mathcal{X}}$-privacy. The geometric mechanism adds integer-valued noise from a symmetric geometric distribution to the output of $f$, ensuring $\epsilon$-$d_{\mathcal{X}}$-privacy.

\begin{definition}
\label{def:geometric_mechanism}
Let $f: \mathcal{X} \rightarrow \mathbb{Z}$ and let $\gamma$ be drawn from a symmetric geometric distribution with the probability mass function:
$G_x = \frac{1-e^{-\epsilon}}{1+e^{-\epsilon}}\cdot e^{-\epsilon \cdot d_{\mathcal{X}}(x,x')}$.
Then, $\mathcal{A}(.) = f(.) + \gamma$ is $\epsilon$-$d_{\mathcal{X}}$-private. 
\end{definition}

%\noindent We prove that the geometric mechanism is $\epsilon$-$d_{\mathcal{X}}$-private in Appendix~\ref{app:proofs}.
\noindent We provide the proof in Appendix~\ref{app:proofs}.

\begin{figure}[h]
    \centering
    \includegraphics[width=0.47\textwidth]{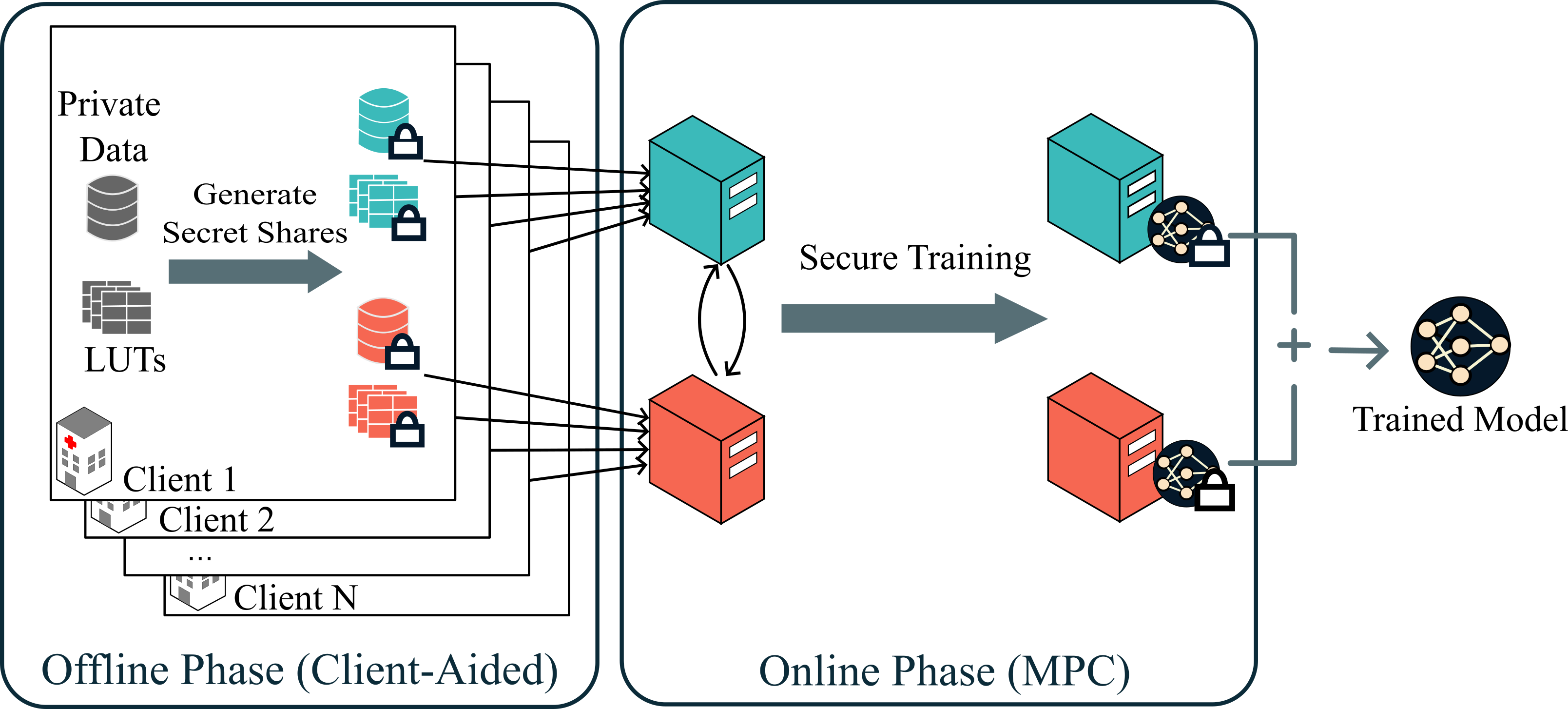}
    \caption{Overview of our PPML setup.}
    \label{fig:overview}
\end{figure}

%% file: security_model.tex
\section{Security Model}
\label{sec:securitymodel}

We consider a setting where clients $C_0,.., C_{n-1}$ want to train ML models on their joint data. The data can be horizontally or vertically partitioned among the clients. The clients outsource the computation to two untrusted but non-colluding servers, $P_0$ and $P_1$. The clients secret share the input data between $P_0$ and $P_1$ that employ secure 2PC techniques to train ML models on this data. Figure \ref{fig:overview} depicts an overview of the setup.

\noindent \textbf{Client-Aided Offline Phase.} We use a client-aided offline phase. Since clients must initially upload their data to the servers, we further mandate them to generate and transmit random numbers and lookup tables, which are later utilized in the online phase. This strategy considerably enhances the efficiency of the offline phase by imposing only a marginal additional load on the clients, who are uninvolved post this phase. An alternative is the server-aided offline phase, where the computing servers use homomorphic encryption and oblivious transfer based techniques to generate required randomness. However, as noted by past work~\cite{secureml}, this method is a significant bottleneck for PPML due to extensive cryptographic tasks. Aiming for practical PPML protocols, we adopt a client-aided offline phase, which provides a significant speed up as seen in Table~\ref{table:training_time_comparison_NN_multi_client}. The only limitation of a client-aided offline phase is that the participating clients can not collude with the computing servers. This ``Client-Aided Setting'' has been formalized and adopted in several past works on PPML~\cite{secureml, outsourcing_mpc_paper, linear_regression_paper}. It is worth noting that each client supplies randomness and lookup tables exclusively for their specific batch of training data. We refer to the client executing the offline phase as a $\csp$.

\subsection{Security Definition}

We consider a semi-honest adversary $\adv$, which adheres to protocol specifications but tries to obtain information about the clients' data. We assume $\adv$ can corrupt at most one of the two servers and any client subset, provided corrupt clients don't collude with servers. The security definition requires that adversary $\adv$ only learns the data of corrupted clients and the final trained model but does not learn the data of any honest client.    

We prove the security of our protocols using the simulation paradigm~\cite{multiparty_composition_paper, universal_composition_paper, simulation_security_paper}. The universal composition framework~\cite{universal_composition_paper} facilitates security arguments for diverse protocol compositions. End-to-end training security follows from the composition of underlying protocols. Security is modeled using two interactions: a real interaction where the servers run a two-party protocol $\Pi$ in the presence of an adversary $\adv$ as defined above and an \textit{environment} $\mathcal{Z}$, and an ideal interaction where the servers send the inputs to a functionality $\mathcal{F}$ that performs the desired computation truthfully.
\begin{definition}
A protocol $\Pi$ securely realizes $\mathcal{F}$ if, for every adversary $\adv$ acting in the real world $\real[\mathcal{Z},\mathcal{A},\Pi, \lambda]$, there is an adversary $\mathcal{S}$ in the ideal world $\ideal[\mathcal{Z},\mathcal{S},\mathcal{F}, \lambda]$, such that the probability of  $\adv$ succeeding in the real world interaction and the probability of $\mathcal{S}$ succeeding in the ideal world interaction, given the security parameter $\lambda$, are related as:

{\footnotesize \noindent $|Pr[\real[\mathcal{Z},\mathcal{A},\Pi, \lambda] = 1] -  Pr[\ideal[\mathcal{Z},\mathcal{S},\mathcal{F}, \lambda] = 1]| = \negl(\lambda)$}

\end{definition}

We introduce two important modifications to the security definition to establish security for $\hawkmulti$ protocol, which permits access pattern leakages. Firstly, we allow some leakage in the ideal world to mirror the information learned by the adversary in the real world by observing the lookup table accesses. The leakage function is a randomized function of the access patterns. Secondly, we impose an extra condition requiring this leakage function to maintain $d_{\mathcal{X}}$-privacy. In the ideal interaction, each party $P_i$ forwards its input to the ideal functionality. The ideal functionality reconstructs the input received $x_1,x_2,..,x_n$, computes the output $f(x_1,x_2,..x_n)$, and sends $\share{f(x_1,x_2,..x_n)}{0} $ to $P_0$ and $\share{f(x_1,x_2,..x_n)}{1} $ to $P_1$. Concurrently, the ideal functionality computes a leakage function $\mathcal{L}$ of the input and sends $\mathcal{L}$ to the simulator $\mathcal{S}$. Our protocols operate in a hybrid world with access to secure, ideal functionalities, which we substitute with secure computation techniques in our application. Per Canetti's classic result~\cite{canetti_paper}, proving security involves treating these as trusted functionality calls.

\begin{definition}
\label{def:hybridmodel}
A protocol $\Pi$ securely realizes $\mathcal{F}$ with leakage $\mathcal{L}$ and $(\lambda, \epsilon)$-security, while making calls to an ideal functionality $\mathcal{G}$, if, $\mathcal{L}$ is $\epsilon$-$d_{\mathcal{X}}$-private, and for every adversary $\adv$ acting in the $\mathcal{G}$-hybrid world $\hybrid[\mathcal{G},\mathcal{Z},\mathcal{A},\Pi, (X_0, X_1), \lambda]$, there is an adversary $\mathcal{S}$ in the ideal world $\ideal[\mathcal{Z},\mathcal{S(L(X))},\mathcal{F}, (X_0, X_1), \lambda]$, such that the probability of  $\adv$ succeeding in the hybrid world interaction and the probability of $\mathcal{S}$ succeeding in the ideal world interaction, on valid inputs $X_0$ and $X_1$, are related as:

{\footnotesize \noindent $|Pr[\hybrid[\mathcal{G}, \mathcal{Z},\mathcal{A},\Pi, (X_0, X_1), \lambda] = 1] - \\  Pr[\ideal[\mathcal{Z},\mathcal{S(L(X))},\mathcal{F}, (X_0, X_1), \lambda] = 1]| = \negl(\lambda)$}

\end{definition}

%% file: protocols.tex
\section{Accelerating Secure Computation with Lookup Tables}
\label{sec:accelerate}

We propose two novel secure computation protocols that employ lookup tables to evaluate a specific univariate function, $y = f(x)$. These protocols enable the secure computation of non-linear activations. Furthermore, we detail how to extend our method to devise specialized protocols for certain multivariate functions. Specifically, we present a protocol for computing the $\softmax$ function using univariate functions. We use the prepossessing model of secure computation discussed in Section~\ref{sec:prelim-secure-computation}.

During the offline phase, for a party $P_i$, a lookup table $L^f_i$ can be constructed, where $f: X \to Y$ is a univariate function and $X$ and $Y$ are the domain and range of $f$, respectively. Each entry in $L^f_i$ is of the form $key: \prf(K, x)$, $value: \share{f(x)}{i}$ for every $x \in X$. Here $\prf(K, \cdot)$ denotes a pseudo-random function with key $K$. In the online phase, $P_i$ holds a secret share $\share{x}{i}$ and seeks to determine $\share{f(x)}{i}$. As $\share{x}{i}$ is a secret share generated in the online phase, it cannot directly serve as a key for the value $\share{f(x)}{i}$. The key for this value must be deterministically derived from $x$. During the online phase, computation parties collaborate to securely compute $\prf(K, x)$ from the shares $\share{x}{i}$ and use it to retrieve $\share{f(x)}{i}$ from the lookup table $L^f_i$. To this end, we propose two protocols $\textsc{Hawk}_{\text{Single}}$ and $\textsc{Hawk}_{\text{Multi}}$. While $\textsc{Hawk}_{\text{Single}}$ provides perfect security, it comes with a higher offline cost and increased storage overhead. In contrast, $\textsc{Hawk}_{\text{Multi}}$ permits table reuse, substantially reducing these overheads, but it does introduce some access pattern leakage, which is proved to be $d_{\mathcal{X}}$-private.

\input{single_use_lookup}

\subsection{$\textsc{Hawk}_{\text{Single}}$: Single Use Lookup Table Protocol}
\label{subsec:single_use_lookup}
In the online phase of $\textsc{Hawk}_{\text{Single}}$, each computation of $f$ consumes an entire table. We generate a distinct table for every lookup during the offline phase. The offline and online phases are detailed in Algorithms~\ref{algorithm:lookup-single-offline}~and~\ref{algorithm:lookup-single-online}, respectively. Here, $c$ ensures each table has distinct keys. For practical utility, the size of the table for $\textsc{Hawk}_{\text{Single}}$ must be minimized. Note that each $\csp$ supplies lookup tables exclusively for their specific batch of training data.

\para{\bf Reducing the Size of Lookup Tables.} We harness recent advancements in low precision machine learning~\cite{limited_precision_training_fixed_pt,HFP8_training}, discussed in Appendix~\ref{app:less_precise_training}, to trim the table size. Neural networks trained with 16-bit fixed-point values retain their accuracy~\cite{limited_precision_training_fixed_pt}. By adopting 16-bit values for activation function inputs and outputs, our lookup tables are constrained to $2^{16}$ entries.

\begin{figure}
    \centering
    
    % 1st Graph
    \begin{subfigure}{0.3\textwidth}
        \begin{tikzpicture}
        \begin{axis}[
            xlabel={Number of Iterations},
            ylabel={Accuracy ($\%$)},
            legend style={at={(1,0.42)}, anchor=north east, font=\fontsize{6}{4}\selectfont, text width=2.3cm, row sep=-2pt},
            legend entries={Plaintext, {Sigmoid<int=3, frac=13>}, {Sigmoid<int=5, frac=11>}, {Sigmoid<int=9, frac=7>}},
            ytick={80,85,90,95,100},
            yticklabels={80,85,90,95,100},
            ymin=75, ymax=100,
        ]
        
        \addplot[color=green,mark=triangle] coordinates {
            (0,96.58) (70, 97.58) (140, 97.85) (210, 98.05) (280, 98.19) (350, 98.31) (420, 98.37) (490, 98.42) (560, 98.55) (630, 98.68) (700, 98.71) 
        };
        
        \addplot[color=blue,mark=square] coordinates {
            (0,98.24) (70,98.55 ) (140, 98.75) (210, 98.84) (280, 98.93) (350, 98.96) (420, 99.01) (490, 99.07) (560,98.97 ) (630, 99.08) (700,99.17 )
        };
        
        \addplot[color=brown,mark=o] coordinates {
            (0,96.58) (70, 97.58) (140, 97.84) (210, 98.04) (280, 98.19) (350, 98.3) (420, 98.37) (490, 98.41) (560, 98.58) (630, 98.7) (700, 98.71)
        };
        
        \addplot[color=red,mark=diamond] coordinates {
            (0,90.2) (70, 90.2) (140, 90.2) (210, 90.2) (280, 90.2) (350, 90.2) (420, 90.33) (490, 91.03) (560, 92.37) (630, 93.1) (700, 93.81)
        };
        
        \end{axis}
        \end{tikzpicture}
        %\caption{MNIST ($|B| = 128$)}
    \end{subfigure}

    \begin{subfigure}{0.3\textwidth}
        \begin{tikzpicture}
        \begin{axis}[
            xlabel={Number of Iterations},
            ylabel={Accuracy ($\%$)},
            legend style={at={(1,0.42)}, anchor=north east, font=\fontsize{6}{4}\selectfont, text width=2.3cm, row sep=-2pt},
            legend entries={Plaintext, {Sigmoid<int=3, frac=13>}, {Sigmoid<int=5, frac=11>}, {Sigmoid<int=9, frac=7>}},
            ytick={10,30,50,70,90},
            yticklabels={10,30,50,70,90},
            ymin=10, ymax=90,
        ]
        
        \addplot[color=green,mark=triangle] coordinates {
            (0, 56) (100, 76) (200, 83) (300, 82) (400, 84) (500, 84) (600, 84) (700, 84) (800, 84) 
        };
        
        \addplot[color=blue,mark=square] coordinates {
            (0, 56) (100, 61) (200, 84) (300, 83) (400, 83) (500, 85) (600, 84) (700, 84) (800, 84)
        };
        
        \addplot[color=brown,mark=o] coordinates {
            (0, 56) (100, 82) (200, 81) (300, 84) (400, 84) (500, 84) (600, 85) (700, 85) (800, 85)
        };
        
        \addplot[color=red,mark=diamond] coordinates {
            (0, 56) (1, 67) (200, 73) (300, 73) (400, 73) (500, 74) (600, 76) (700, 76) (800, 77)
        };
        
        \end{axis}
        \end{tikzpicture}
        %\caption{Arcene ($|B| = 100$)}
    \end{subfigure}

    \caption{a) MNIST ($\textbf{|B| = 128}$) \hfill b) Arcene ($\textbf{|B| = 100}$)\\
         Accuracy comparison of plaintext training with our PPLR protocol considering different bit representations for the Sigmoid function. %($\alpha = 2^{-5}$)
         }
    \label{fig:reduced_precision_training}
\end{figure}
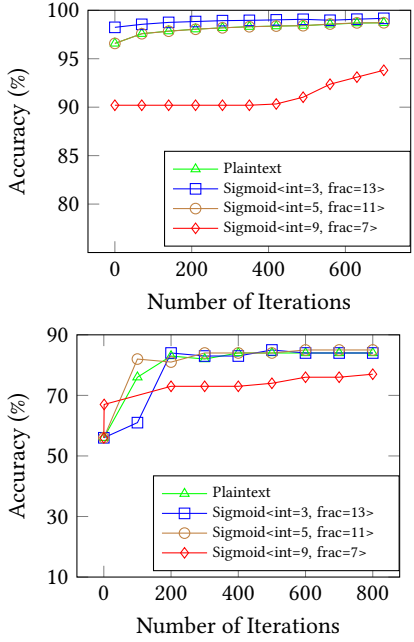

Our experiments on the MNIST dataset (classifying as ``0'' or ``not 0'') and the Arcene dataset (distinguishing ``cancer'' from ``normal'') validate that 16-bit fixed-point numbers suffice for activation functions without compromising model accuracy. Figure~\ref{fig:reduced_precision_training} shows accuracy results for privacy-preserving logistic regression (PPLR) using our protocols, which employ 16-bit fixed-point representation for $\sigmoid$, and provides a comparison with plaintext training using 32-bit floating-point numbers. We observe that our protocol retains the model accuracy. Specifically, allocating 11 to 13 bits for the fractional component and the remaining for the integer part yields accuracy nearly identical to plaintext training.

\para{\bf Optimizing Lookups for Derivatives.} In certain cases, we can compute the derivative from the activation function or vice versa. For instance, $\dsigmoid(x) = \sigmoid(x)(1-\sigmoid(x))$ and $\relu(x) = x \drelu(x))$, where $\dsigmoid$ and $\drelu$ represent the derivative of $ \sigmoid$ and $ \relu$ respectively. Our protocol, which leverages this relationship for $\relu$ and $\drelu$, is detailed in Algorithm~\ref{algorithm:relu}. Once the parties have $\drelu(x)$, they can use a secure multiplication to compute $\relu(x)$. This substantially cuts down the need for additional lookup tables. While the concept of computing $\relu(x)$ from $\drelu(x)$ is not novel~\cite{securenn}, our contribution lies in harnessing this relationship to halve the lookup tables needed for neural networks, thereby streamlining the offline phase.

\para{\bf An Alternate Representation for $ \mathsf{\textbf{DReLU}}$.}
We present an alternate representation of the $\drelu$ activation, aiming to reduce the size of its associated lookup table substantially. Note that $\drelu(x)$ = 1 if $x > 0$ and 0 otherwise. Our key observation is that, unlike other activation functions, $\drelu$ does not require precise numbers. For instance, given $x_1 = 2$ and $x_2 = 2.57$, both yield $\drelu(x_1)=\drelu(x_2)=1$. This suggests that the fractional component of the number can be disregarded without affecting the computation of $\drelu$. However, for $0 < x < 1$, eliminating the fractional component results in $x=0$, leading to incorrect computation of $\drelu(x)=0$. To address this, we propose the following representation for $\drelu$ for fixed point numbers: 
$\drelu(x) = 1$ iff $\floor{x+(2^{l_{d}}-1)} \geq 1$ and $0$ otherwise, where $l_{d}$ represents bit count for the decimal part and $\floor{.}$ signifies truncation by $l_{d}$ bits. This novel representation can substantially reduce the $\drelu$ lookup table entries, allowing us to drop fractional bits and only retain integer bits. For instance, if $l_{d}$ bits represent the fractional segment of a number and $l_{i}$ bits correspond to the integer segment, a lookup table requires $2^{l_{d} + l_{i}}$ entries. With our proposed representation, only $2^{l_{i}}$ entries are needed, reducing the lookup table size for $\drelu$ by a factor of $2^{l_{d}}$. In our context, this factor is $2^{13}$. As highlighted in previous work~\cite{secureml}, each participant $P_i$ can locally truncate their share of $x$. However, as this truncation introduces a small error (refer to Theorem 1 in~\cite{secureml} for details), we opt to retain one fractional bit instead of truncating all of them.

\para{\bf Addressing Storage Overhead.} Despite our optimizations, the storage demands of lookup tables render $\textsc{Hawk}_{\text{Single}}$ suitable only for logistic regression and secure inference but not for neural network training, as shown in Table~\ref{table:training_time_comparison_NN_multi_client}. To address this, we introduce $\textsc{Hawk}_{\text{Multi}}$, which permits multiple reuses of a single lookup table. While this approach introduces some leakage regarding access patterns, we prove the leakage is bounded by $\epsilon$-$d_{\mathcal{X}}$-privacy.

% While this approach introduces some leakage on the accessed indexes of an encrypted lookup table, we prove it preserves $\epsilon$-$d_{\mathcal{X}}$-privacy.

\input{share_convert}
\input{multi-use-lookup-new}

\subsection{$\textsc{Hawk}_{\text{Multi}}$: Multi-Use Lookup Table Protocol}
Algorithms~\ref{algorithm:lookup-multi-offline} and~\ref{algorithm:lookup-multi-online} detail the offline and online phases of our $\textsc{Hawk}_{\text{Multi}}$ protocol, respectively. The core concept of the $\textsc{Hawk}_{\text{Multi}}$ protocol is the efficient reuse of a lookup table for multiple lookups. However, naively reusing a lookup table poses several challenges, as elaborated below.

\para{\bf Lookup Table Reuse.} Directly reusing a lookup table to compute $f(x)$ for identical $x$ values multiple times risks leaking the frequency distribution of $x$. While parties might not directly learn $x$, they can deduce a histogram representing each key's access frequency in the table. As demonstrated by Naveed et al.~\cite{naveed2015inference}, this leakage may reveal plaintext when combined with auxiliary data.

Rather than entirely eliminating this access pattern leakage as in ORAM which has an inherent overhead of $\Omega(log(n))$ \cite{goldreichSoftwareProtectionSimulation1996, boyle2016there, larsen2018yes}, our $\textsc{Hawk}_{\text{Multi}}$ protocol employs $d_{\mathcal{X}}$-privacy to limit it. This approach offers a balance between efficiency and privacy, precisely quantifying the leaked information. We analyze this leakage in detail in Section~\ref{sec:leakageanalysis}. With the $\textsc{Hawk}_{\text{Multi}}$ protocol, a lookup table remains reusable until the allocated privacy budget $\epsilon_{T}$ is depleted. Once exhausted, the table is discarded in favor of a new one.

\para{\bf Secure Deterministic Key Lookup.}  The objective is to securely compute a deterministic key $\prf(K, x)$ from the shares $\share{x}{i}$ and use it to retrieve the value $\share{f(x)}{i}$ from the lookup table $L^f_i$. In the $\textsc{Hawk}_{\text{Single}}$, this is achieved by using the same pseudorandom pad $H(k_i || c)$  to obscure $x$ for every key-value pair in a lookup table. This approach remains secure as each table is utilized once per activation function computation, and the pad is never reused. 

In the $\textsc{Hawk}_{\text{Multi}}$ protocol, we utilize the hardness of the \textit{elliptic curve discrete logarithm problem (ECDLP)}~\cite{ECDLP} to securely compute $\prf(K, x)$. ECDLP, foundational to elliptic curve cryptography (ECC), posits that given elliptic curve points $P, Q \in E(\mathbb{F}_q)$, determining $a$ such that $Q = aP$ is computationally challenging. ECC is integral to $\textsc{Hawk}_{\text{Multi}}$ for two additional properties, homomorphism and efficiency. ECC naturally supports homomorphic operations such as $xG = \share{x}{0}^{N}G + \share{x}{1}^{N}G$, where $N$ denotes the elliptic curve order and $G$ is the curve generator. We leverage this homomorphism to securely compute $\prf(K, x)$. Furthermore, compared to RSA, ECC requires $12\times$ smaller keys for 128 bits of security, leading to faster computations and reduced communication~\cite{ecc_vs_rsa}.

We iteratively detail the protocol design, starting with the core concept and refining it to the finalized protocol. Suppose parties $P_0$ and $P_1$ possess shares $\share{x}{0}^N$ and $\share{x}{1}^N$ of $x$. They also have secret random keys $k_0$ and $k_1$, respectively. In the subsequent section, we present a protocol to convert a share $\share{x}{i}$ of $x$ from $\share{x}{i}^n$ to $\share{x}{i}^N$. $P_0$ computes $k_{0}(\share{x}{0}^{N}G)$ and forwards it to $P_1$ which then computes $k_{1}k_{0}( \share{x}{0}^{N}G)$. Similarly, $P_1$ sends $k_{1}(\share{x}{1}^{N}G)$ to $P_0$ which computes $k_{0}k_{1}( \share{x}{1}^{N}G)$. Due to the ECDL assumption, $k_{i}(\share{x}{i}^{N}G)$ does not leak information about $\share{x}{i}^N$ or $k_i$. Subsequently, $P_1$ sends $k_{1}k_{0}( \share{x}{0}^{N}G)$ to $P_0$ and $P_0$ sends $k_{0}k_{1}( \share{x}{1}^{N}G)$ to $P_1$. Each party $P_i$ can then compute $\kappa = k_{0}k_{1}( xG) \equiv  k_{0}k_{1}( \share{x}{1}^{N}G) + k_{0}k_{1}( \share{x}{0}^{N}G)$. This allows $P_0$ and $P_1$ to securely compute $\prf(K, x)$, where $K = k_0 k_1$ which can be used as a key for table lookup.

\sloppy \para{\bf Converting Additive Shares To A Larger Group.}
For the $\textsc{Hawk}_{\text{Multi}}$ protocol, we start by converting additive shares from $\mathbb{Z}_n$ to $\mathbb{Z}_N$, where $N > n$ and $N$ is the order of the elliptic curve $E$. Algorithm~\ref{algorithm:share-convert} presents a one-round protocol for share conversion based on ~\cite{share_conversion_paper}. The protocol requires shares of a random $r$ in both $\mathbb{Z}_n$ and $\mathbb{Z}_N$, which can be precomputed in the offline phase.

\para{\bf Mitigating Malleability.} Given $k_{0}k_{1}(xG)$, it is possible to compute $k_{0}k_{1}\cdot$ $(axG) \equiv a \times (k_{0}k_{1}(xG))$ for any arbitrary $a$. While this does not directly reveal $x$, it might leak crucial information in our protocol when multiple lookups are performed on a table. A malicious party could compute $k_{0}k_{1}(axG)$ for various $a$ values and cross-check all past and future lookups $k_{0}k_{1}(bG)$ to determine if any $k_{0}k_{1}(axG) = k_{0}k_{1}(bG)$, thereby inferring if $b$ is a multiple of $x$. To counter this, we use $k_{0}k_{1}(xG + s_{0}G + s_{1}G)$ as the lookup table key. Here $k_{0}$ and $s_{0}$ are random keys known only to $P_0$, while $k_{1}$ and $s_{1}$ are known only to $P_1$. Having these two sets of keys ensures no party $P_i$ can independently compute $k_{0}k_{1}(axG + s_{0}G + s_{1}G)$. Furthermore, instead of directly using $s_0$ and $s_1$, the parties use distinct $s_0^{c}$ and $s_1^{c}$ for each table where $c$ denotes the lookup table index and $s^c_i = H(s_i||c)$. This guarantees that the lookup key differs for each $x$ across different lookup tables.

\para{\bf Reducing Number of Elliptic Curve Scalar Multiplications.}
%\noindent We present two optimizations to significantly reduce elliptic curve (EC) multiplications in $\textsc{Hawk}_{\text{Multi}}$ protocol. The first optimization cuts down EC multiplications from $m \times 2^{18}$ to $m+1$ and point additions from $m \times 2^{17}$ to $m \times 2^{16}$ in Algorithm~\ref{algorithm:lookup-multi-offline}. The second optimization reduces three EC multiplications and one addition in steps 5 and 6 of Algorithm~\ref{algorithm:lookup-multi-online} to a single scalar multiplication. We refer the reader to Appendix~\ref{app:optimizations} for details about these optimizations.

\noindent \textit{Optimization 1.} During the offline phase, a $\csp$ must generate $m$ tables, each of size $2^{16}$. A significant computational bottleneck in our $\textsc{Hawk}_{\text{Multi}}$ protocol arises when computing elliptic curve scalar multiplications in step 2.3a of Algorithm~\ref{algorithm:lookup-multi-offline}. This step requires four scalar multiplications and two point additions on the elliptic curve. This translates to a daunting $m \times 2^{18}$ scalar multiplications and $m \times 2^{17}$ point additions in total. We introduce an optimization to reduce these numbers drastically: scalar multiplications are cut down to $m+1$, and point additions to $m \times 2^{16}$. Specifically, the $\csp$ initially calculates $(k_0k_1)\cdot G$. For a table $L^{f,c}$, it then computes $(k_0k_1)(s^c_0 + s^c_1)\cdot G$. The key for $x=0$ is set as $\prf(K, 0)=(k_0k_1)(s^c_0 + s^c_1)\cdot G$. All subsequent keys are iteratively determined as $\prf(K, i) = \prf(K, i-1)+(k_0k_1)\cdot G$, requiring only a single point addition. 

\noindent \textit{Optimization 2.} In Algorithm~\ref{algorithm:lookup-multi-online}, steps 5 and 6, if executed directly, demand three scalar multiplications and one point addition for each party on the elliptic curve. We suggest an optimization to bring this down to a single scalar multiplication per party. Each party $P_i$ first evaluates $k_i(\share{x}{i}^{N}+s^c_i)$, and subsequently computes $k_i(\share{x}{i}^{N}+s^c_i)\cdot G$, which requires just one scalar multiplication.

\para{\bf Ensuring Differentially Private Access Patterns.}
While our $\textsc{Hawk}_{\text{Multi}}$ protocol conceals the input $x$ during lookups, repeated lookups on the same table might inadvertently reveal a histogram of the frequency of each pseudorandom key's access in the table. To address this, our $\textsc{Hawk}_{\text{Multi}}$ protocol uses $d_{\mathcal{X}}$-privacy to limit this leakage. Rather than using their actual input share $\share{x}{i}$, each party computes a perturbed input share $\share{\hat{x}}{i}$ by incorporating $d_{\mathcal{X}}$-private noise $\share{\gamma} {i}$ as outlined in step 1 of Algorithm~\ref{algorithm:lookup-multi-online}. This ensures that the parties only learn noisy access patterns, maintaining $\epsilon$-$d_{\mathcal{X}}$-privacy. The shares of noise $\gamma$ are generated in the offline phase.

\subsection{Computing Activation Functions}
\label{subsec:Act-Func}

We now describe how we compute different activation functions using our protocols. We use $\Pi^\online _\lookup$ to denote either $\Pi^\online _{\textsc{Hawk}_{\text{Single}}}$ or  $\Pi^\online _{\textsc{Hawk}_{\text{Multi}}}$.

\input{relu_derivative}

\input{relu}

\para{\bf ReLU and DReLU.}
Algorithm~\ref{algorithm:drelu} and ~\ref{algorithm:relu} describe our protocols for computing $\drelu$ and $\relu$ respectively. We use the property: $\relu(x) = x \drelu(x)$, to compute $\relu(x)$ from $\drelu(x)$. 

\input{sigmoid}

\para{\bf Sigmoid.}
Algorithm~\ref{algorithm:sigmoid} describes our protocol to compute $\sigmoid(x)$. It is worth noting that the prior works~\cite{secureml, securenn} do not provide any method to compute $\sigmoid$ accurately.

\para{\bf Softmax.}
Algorithm~\ref{algorithm:softmax} describes our protocol for computing $\softmax$. As $\softmax$ involves $d_l$ inputs, using our protocols naively requires $2^{n \times d_{l}}$ entries in the lookup table given $x_j$ is an $n$-bit number, which is impractical. We introduce a method that substantially reduces the lookup table size, albeit with a slight increase in online computation. Rather than directly computing $\softmax$ using a single lookup table, we employ two distinct tables: one for the exponential function, $\expo(x) = e^x$, and another for the inverse, $\inverse(x) = \frac{1}{x}$. Existing literature lacks a precise method for $\softmax$ computation, resorting to non-standard functions~\cite{secureml, securenn}, which leads to accuracy degradation ( see Figure~\ref{fig:convergence_comparison_NN} ). 
%Furthermore, reliance on non-standard functions can cause accuracy to plummet after a certain number of epochs, as depicted in Figure~\ref{fig:convergence_comparison_NN}.

\input{softmax}

\subsection{Rounds and Communication Complexity}
\label{subsec:rounds_and_communication}

Table~\ref{table:rounds_and_complexity} details the rounds and communication complexities of sub-protocols employed in training both logistic regression models and neural networks. For the $\textsc{Hawk}_{\text{Single}}$ protocol, with $l$-bit numbers ($l=64$), the associated protocols: $\Pi_{\lookup}$, $\Pi_{\sigmoid}$ and, $\Pi_{\drelu}$, demand a single communication round with a complexity of $l$ bits. The $\Pi_{\relu}$  requires an extra round due to an additional secure multiplication following $\drelu$. Meanwhile, $\Pi_{\softmax}$ requires three rounds, attributed to the computations of the exponential function, the inverse function, and a secure multiplication. Here, $d_{l}$ denotes the number of neurons in the output layer.

For the $\textsc{Hawk}_{\text{Multi}}$ protocol, we adopt numbers of $l' = 256$ bits. The associated protocols: $\Pi_{\lookup}$, $\Pi_{\sigmoid}$, and $\Pi_{\drelu}$, require three communication rounds with a complexity of $l+2l'$. The $\textsc{Hawk}_{\text{Multi}}$ based protocol $\Pi_{\softmax}$ requires a total of seven rounds.

The efficiency improvements in training PPML models in our case, relative to SecureML, stem from the fast computation of non-linear activations such as $\relu$, $\sigmoid$, $\softmax$, and their derivatives. Past works~\cite{secureml, mini_onn, gazelle} leveraged Yao's garbled circuits for these computations. Transitioning secret shares between arithmetic and Yao sharing introduces an overhead of 7$l\kappa$ + ($l^2$ + $l$)$/2$~\cite{ABY_paper}. 

%% file: single_use_lookup.tex
\begin{algorithm}[t]
    \caption{Table Generation: $\Pi_{\textsc{Hawk}_{\text{Single}}(\csp)}^\offline$}
    \label{algorithm:lookup-single-offline}

    \begin{enumerate}

    \Require Total number of lookup tables $m$, function $f: X \to Y$, and a cryptographic hash function $H$.

    \hspace{-3.6em} \textbf{Output:} $P_i$ gets lookup tables $L^{f,c}_{i}$ for $0 \leq c < m$.

    \item $\csp$ sends a random key $k_i$ to $P_i$.

    \item For $0 \leq c \leq m-1$:
        \begin{enumerate}
         \item For all $x \in X$:
             \begin{enumerate}[leftmargin=*]
    
            \item $\csp$ computes additive secret shares $\langle {y} \rangle_0$ and $\langle {y} \rangle_1$ of $y = f(x)$.
            
             \item $\csp$ adds the $key: H(x+H(k_1||c))$, $value: \langle {y} \rangle_0$ pair in $L^{f,c}_0$.
             
             \item $\csp$ adds the $key: H(x+H(k_0||c))$, $value: \langle {y} \rangle_1$  pair in $L^{f,c}_1$.
             
             \end{enumerate}
             
             \item $\csp$ randomly permutes all key value pairs in $L^{f,c}_0$ and $L^{f,c}_1$.

        \end{enumerate}
        
        \item  $\csp$ sends all $L^{f,c}_0$ for $0 \leq c < m$ to $P_0$.
        \item  $\csp$ sends all $L^{f,c}_1$ for $0 \leq c < m$ to $P_1$.
     
    \end{enumerate}
\end{algorithm}

\begin{algorithm}[t]
    \caption{Query Table: $\Pi_{\textsc{Hawk}_{\text{Single}}(P_0, P_1)}^\online$}
    \label{algorithm:lookup-single-online}
    
    \begin{algorithmic}[1]

    \Require For $i \in \{0,1\}$,  $P_i$ holds key $k_i$,  $\langle {x} \rangle_i$, and a cryptographic hash function $H$.

    \Ensure $P_i$ gets secret share $\langle {y} \rangle_i$ of $y = f(x)$.

    \Statex \hspace{-2em} \textbf{Initialization:} $P_i$ receives lookup tables $L^{f,c}_{i}$ for all $0 \leq c < m$ from $\csp$. $P_0$ and $P_1$ initialize $c = 0$.
        
    \item $P_i$ computes $\share{x}{i} + H(k_i || c)$ and sends it to $P_{1-i}$.

    \item $P_i$ computes $H(x + H(k_{1-i} || c)) = H(\share{x}{i} + \share{x}{{1-i}} + H(k_{1-i} || c))$.
    
    \item $P_i$ retrieves the value: $\langle {y} \rangle_i$ corresponding to the key: $H(x + H(k_{1-i} ||c))$ from $L^{f,c}_{i}$.

    \item $P_0$ and $P_1$ increment $c = c + 1$.
    
    \item $P_i$ outputs $\langle {y} \rangle_i$.

    \end{algorithmic}
\end{algorithm}

%% file: share_convert.tex
\begin{algorithm}

    \caption{$\Pi_{\mathsf{SC}(P_0, P_1)} ^ {\online}$: Share Conversion Protocol}
    \label{algorithm:share-convert}

    \begin{algorithmic}[1]
    \Require For $i \in \{0,1\}$, $P_i$ knows $n$ and $N$ where $N>n$. $P_i$ holds $\share{x}{i}^n$ and shares of a random $r$; $\share{r}{i}^n$ and $\share{r}{i}^N$.
    \Ensure $P_i$ gets fresh share $\share{x}{i}^{N}$ of $x$.
    \item $P_i$ computes $\share{x}{i}^n - \share{r}{i}^n$ and sends it to $P_{1-i}$ 
    \item Each party reconstructs $z = x - r$. 
     \item $P_0$ outputs $\share{x}{0}^{N} = z + \share{r}{0}^{N}$ and $P_1$ outputs $\share{x}{1}^{N} = \share{r}{1}^{N}$  as their share of $x$ in \( \mathbb{Z}_N \)
    \end{algorithmic}
\end{algorithm}

%% file: multi-use-lookup-new.tex
\begin{algorithm}
\small
    \caption{Table Generation: $\Pi_{\textsc{Hawk}_{\text{Multi}}(\csp)}^\offline$}
    \label{algorithm:lookup-multi-offline}
    \begin{enumerate}
    \Require A function $f: X \to Y$, total tables required $m$, cryptographic hash function $H$, field modulus $n$, order of an elliptic curve $N$, generator of the curve $G$.
    
    \hspace{-3.8em} \textbf{Output:} For $i \in \{0,1\}$, $P_i$ gets lookup tables $L^{f,c}_{i}$ for $0 \leq c < m$.
        
     \item $\csp$ generates random keys $k_i$ and $s_i\mod{N}$ and sends to $P_i$.          

     \item For $0 \leq c \leq m - 1$: 
    \begin{enumerate}
        \item $\csp$ computes random $s^c_0 = H(s_0||c)\mod{N}$.
        \item $\csp$ computes random $s^c_1 = H(s_1||c)\mod{N}$.
        \item For all $x \in X$: 
        \begin{enumerate}[leftmargin=*]
            \item $\csp$ computes $\kappa = k_0 k_1 \times (xG + s^c_{0}G+s^c_{1}G )$.
            \item $\csp$ computes additive secret shares $\langle {y} \rangle_0$ and $\langle {y} \rangle_1$ of $y = f(x)$.
            \item $\csp$ adds $key: H(\kappa), value: \langle {y} \rangle_0$ in $L^{f}_0$.
            \item $\csp$ adds $key: H(\kappa), value: \langle {y} \rangle_1$ in $L^{f}_1$.
        \end{enumerate}
        \item $\csp$ randomly permutes all key-value pairs in $L^{f,c}_0$ and  $L^{f,c}_1$.
    \end{enumerate}
        
     \item  $\csp$ sends all $L^{f,c}_0$ for $0 \leq c < m$ to $P_0$.
     \item $\csp$ sends all $L^{f,c}_1$ for $0 \leq c < m$ to $P_1$.
    \end{enumerate}
\end{algorithm}

\begin{algorithm}
%\linespread{0.1}\selectfont
\small
    \caption{Query Table: $\Pi_{\textsc{Hawk}_{\text{Multi}}(P_0, P_1)}^\online$}
    \label{algorithm:lookup-multi-online}
     
    \begin{algorithmic}[1]

        \Require  For $i \in \{0,1\}$, 
     $P_i$ holds random keys $k_i$ and $s_i$,
     $\langle {x} \rangle_{i}^{n}$ , 
     a cryptographic hash function $H$, $\share{r}{i}^{n}$ and $\share{r}{i}^{N}$, order of an elliptic curve $N$, share of $d_{\mathcal{X}}$-private noise $\share{\gamma}{i}^{n}$.
   \Ensure $P_i$ gets secret share $\langle {y} \rangle_i$ of $y = f(x)$.
   \Statex \hspace{-2em} \textbf{Initialization:} $P_i$ receives lookup tables $L^{f,c}_{i}$ for all $0 \leq c < m$ from $\csp$, 
    $P_i$ initializes $c = 0$. $P_i$ receives per query privacy budget $\epsilon$ and total privacy budget $\epsilon_{T}$. $P_i$ initializes $\epsilon_{r}=\epsilon_{T}$. 

        \item $P_i$ computes noisy input $\share{\hat{x}}{i}^{n} = \share{x}{i}^{n} + \share{\gamma}{i}^{n}$
                        
        \item $P_i$ executes $\Pi_{SC(P_0, P_1)} ^ {\online}$ with inputs $\share{\hat{x}}{i}^{n}$, $\share{r}{i}^{n}$ and $\share{r}{i}^{N}$ and receives output $\share{\hat{x}}{i}^{N}$.

        \item $P_0$ computes random value
        $s^c_0 = H(s_0||c)\mod N$.
        
	    \item
        $P_1$ computes random value
        $s^c_1 = H(s_1||c)\mod N$.
        
        \item $P_0$ computes 
	    $k_{0}( \share{\hat{x}}{0}^{N}G + s^{c}_{0}G )$ 
	    and sends it to $P_1$. 
	    
	    \item $P_1$ computes 
	    $k_{1}( \share{\hat{x}}{1}^{N}G + s^{c}_{1}G )$ 
	    and sends it to $P_0$. 
	    
	    \item $P_0$ computes 
	    $k_{0}k_{1}( \share{\hat{x}}{1}^{N}G + s^{c}_{1}G )$ 
	    and sends it to $P_1$. 
	    
	    \item $P_1$ computes 
	    $k_{1}k_{0}( \share{\hat{x}}{0}^{N}G + s^{c}_{0}G )$ 
	    and sends it to $P_0$.
        
	     \item $P_i$ computes 
	    $\kappa = k_{0}k_{1}( \hat{x}G + s^{c}_{0}G+ s^{c}_{1}G ) \equiv 
	    k_{0}k_{1}( \share{\hat{x}}{1}^{N}G + s^{c}_{1}G ) + k_{0}k_{1}( \share{\hat{x}}{0}^{N}G + s^{c}_{0}G )$.
	    
		\item 
		$P_i$ retrieves the value: $\share{y}{i}$ corresponding to the key: $H(\kappa)$ from $L^{f,c}_i$, and outputs $\share{y}{i}$.

    \item $P_i$ computes $\epsilon_{r} = \epsilon_{r} - \epsilon$.
        
        \item If $\epsilon_{r} == 0$ :
            \begin{enumerate}[label=\theenumi.\arabic*]
            \item $P_i$ increments $c = c + 1$.
            \item $P_i$ sets $\epsilon_{r} = \epsilon_{T}$
            \end{enumerate}
        
    \end{algorithmic}
\end{algorithm}

%% file: relu_derivative.tex
\begin{algorithm}[ht]
    \caption{DReLU: $\Pi_{\drelu(P_0, P_1)}$ }
    \label{algorithm:drelu}
    
    \begin{algorithmic}[1]
     \Require $P_i$ holds $\langle {x} \rangle_i$, lookup table $L^{\drelu}_i$.
    \Ensure $P_i$ gets $\langle z \rangle_i$ = $\langle \drelu(x) \rangle_i$ 
     \item  $P_i$ calls $\Pi_{\lookup(P_{i}, P_{1-i})}^\online$ with inputs $(\floor{\langle {x} \rangle_i+i\times(2^{l_{d}}-1)}$, $L^{\drelu}_i)$
     \item $P_i$ learns $\langle z \rangle_i$ = $\langle \drelu(x) \rangle_i$.
     
    \end{algorithmic}
    \end{algorithm}

%% file: relu.tex
\begin{algorithm}[ht]
    \caption{ReLU: $\Pi_{\relu(P_0, P_1)}$}
    \label{algorithm:relu}
    \begin{algorithmic}[1]
    \Require $P_i$ holds $\langle {x} \rangle_i$, $\langle \drelu(x) \rangle_i$.
    \Ensure $P_i$ gets $\langle z \rangle_i$ = $\langle \relu(x) \rangle_i$
     %\item For $i \in \{0, 1\}$, $P_i$ calls $\Pi_{\drelu(P_{i}, P_{1-i})}$ with inputs $(\langle {x} \rangle_i, L^{\drelu}_i)$ and learns $\langle \drelu(x) \rangle_i$.
     \item For $i \in \{0, 1\}$, $P_i$ calls $\Pi_{\secmul(P_{i}, P_{1-i})}$ having inputs $(\langle \drelu(x) \rangle_i, \langle x \rangle_i)$ and learns $\langle z \rangle_i = \langle x \drelu(x) \rangle_i$. Here $\Pi_{\secmul}$ denotes secure multiplication using Beaver's triples.
     \item  $P_i$ outputs $\langle z \rangle_i$ = $\langle \relu(x)\rangle_i = \langle x \drelu(x) \rangle_i$.
     
    \end{algorithmic}
    \end{algorithm}

%% file: sigmoid.tex
\begin{algorithm}[ht]
    \caption{Sigmoid: $\Pi_{\sigmoid(P_0, P_1)}$}
    \label{algorithm:sigmoid}
    
   % $\langle Sgm(x) \rangle_0$ and $\langle Sgm(x)\rangle_1$.
    
    \begin{algorithmic}[1]
    \Require $P_i$ holds $\langle {x} \rangle_i$, lookup table $L^{\sigmoid}_i$. 
    \Ensure $P_i$ gets $\langle z \rangle_i$ =  $\share{\sigmoid(x)}{i}$.
     \item \sloppy For $i \in \{0, 1\}$, $P_i$ calls $\Pi_{\lookup(P_{i}, P_{1-i})}^\online$ with inputs $(\langle {x} \rangle_i, L^{\sigmoid}_i)$
     
     \item $P_i$ learns $\langle z \rangle_i$ = $\langle \sigmoid(x) \rangle_i$.
     
    \end{algorithmic}
    \end{algorithm}

%% file: softmax.tex
\begin{algorithm}[t]
    \caption{Softmax: $\Pi_{\softmax(P_0, P_1)}$}
    \label{algorithm:softmax}
    
    \begin{algorithmic}[1]

     \Require For $i \in \{0, 1\}$ and $j \in \{1, .., d_l\}$, $P_i$ holds $\langle {x_j} \rangle_i$, lookup tables $L^{\exp}_i$ and $L^{\inverse}_i$.
    \Ensure For neuron $j \in \{1, .., d_l\}$, $P_i$ gets $\langle \softmax(x_j) \rangle_i$.
    
     \item For $j \in \{1, .., d_l\}$, $P_i$ calls $\Pi^\online_{\lookup(P_{i}, P_{1-i})}$ with inputs $(\langle {x_j} \rangle_i, L^{\exp}_i)$ and learns $\langle e^{x_j} \rangle_i$.
     
     \item $P_i$ locally computes $\langle s \rangle_i = \sum_{j=1}^{d_l} \langle e^{x_j} \rangle_i$ as the sum of the exponential terms.
     
     \item $P_i$ calls $\Pi^\online _{\lookup(P_i, P_{1-i})}$ with inputs $(\langle {s} \rangle_i, L^{\inverse}_i)$ and learns $\langle \frac{1}{s} \rangle_i$ as the share of the inverse of the sum $s$.
     
     \item For $j \in \{1, .., d_l\}$, $P_i$ runs $\Pi_{\secmul(P_{i}, P_{1-i})}$ with inputs $(\langle e^{x_j} \rangle_i, \langle \frac{1}{s}  \rangle_i)$ and learns $\langle \softmax(x_j) \rangle_i$.
     
    \end{algorithmic}
\end{algorithm}

%% file: security_analysis.tex
\section{Security Analysis}
\label{sec:proofs}

\sloppy Our privacy-preserving training and inference protocols primarily modify SecureML's activation function computation, replacing it with our $\textsc{Hawk}_{\text{Single}}$ or $\textsc{Hawk}_{\text{Multi}}$ based protocols: $\Pi_{\sigmoid}$, $\Pi_{\relu}$, $\Pi_{\drelu}$, and $\Pi_{\softmax}$. Each protocol either exclusively employs $\Pi_{\textsc{Hawk}_{\text{Single}}}^\online$ or $\Pi_{\textsc{Hawk}_{\text{Multi}}}^\online$, or combines them with secure multiplication. Consequently, our security analysis focuses on the $\Pi_{\textsc{Hawk}_{\text{Single}}}^\online$ and $\Pi_{\textsc{Hawk}_{\text{Multi}}}^\online$ protocols, given their composability. The overall security of our protocols follows from SecureML's subprotocols composed with our $\Pi_{\textsc{Hawk}_{\text{Single}}}^\online$ or $\Pi_{\textsc{Hawk}_{\text{Multi}}}^\online$ protocols. 

\input{rounds_and_communication_merged}

\begin{figure}[ht]
\centering
\noindent\fbox{
\parbox{0.45\textwidth}{
\textbf{Functionality} $\mathcal{F}_{\lookupsingle}$

\textbf{Parameters:} Servers $P_0, P_1$

\textbf{Data:} On input $\share{x}{0}$ and $\share{x}{1}$, store $\share{x}{0}$ and $\share{x}{1}$ internally.

\textbf{Computation:} On input $f$ from $P_0$ or $P_1$, reconstruct $x = \share{x}{0} + \share{x}{1} \pmod{n}$, compute $y = f(x)$, create additive shares $\share{y}{0}$ and $\share{y}{1}$, and send $\share{y}{0}$ to $P_0$ and $\share{y}{1}$ to $P_1$.
}}
\caption{Ideal Functionality $\mathcal{F}_{\lookupsingle}$}

\end{figure}

\begin{figure}[ht]
\centering
\noindent\fbox{
\parbox{0.45\textwidth}{
\textbf{Functionality} $\mathcal{F}_{\mathsf{SC}}$

\textbf{Parameters:} Servers $P_0, P_1$

\textbf{Data:} On input $\share{x}{0}^n$ and $\share{x}{1}^n$, store $\share{x}{0}^n$ and $\share{x}{1}^n$ internally.

\textbf{Computation:} On input $N>n$ from $P_0$ or $P_1$, reconstruct $x = \share{x}{0}^n + \share{x}{1}^n \pmod{n}$, create random additive shares $\share{x}{0}^N$ and $\share{x}{1}^N$ of $x$, and send $\share{x}{0}^N$ to $P_0$ and $\share{x}{1}^N$ to $P_1$.
}}
\caption{Ideal Functionality $\mathcal{F}_{\mathsf{SC}}$}

\end{figure}

\begin{figure}[ht]
\centering
\noindent\fbox{
\parbox{0.45\textwidth}{
\textbf{Functionality} $\mathcal{F}_{\lookupmulti}$

\textbf{Parameters:} Servers $P_0, P_1$

\textbf{Data:} On inputs $\share{x}{0}$ and $\share{x}{1}$, store internally.

\textbf{Computation:} 

\begin{enumerate}[leftmargin=*]

\item  Reconstruct $x = \share{x}{0} + \share{x}{1} \pmod{n}$.

\item Compute $\hat{x} = x + \gamma \pmod{n}$, with $\gamma$ drawn using geometric mechanism in Definition~\ref{def:geometric_mechanism}.

\item On input $f$ from $P_0$ or $P_1$, compute  $y = f(\hat{x})$.

\item Create additive shares $\share{y}{0}$ and $\share{y}{1}$ of $y$.

%\item Compute $\kappa =  k_{0}k_{1}( \hat{x}G + s_{0}G+ s_{1}G) $.

%\item Let $\mathcal{L} = \kappa$, where $\mathcal{L}$ is the leakage function.

%\item Send $\share{y}{0}$ and $\mathcal{L}$ to $P_0$ and $\share{y}{1}$ and $\mathcal{L}$ to $P_1$.

\item Send $\share{y}{0}$ to $P_0$ and $\share{y}{1}$ to $P_1$.

\end{enumerate}
}}
\caption{Ideal Functionality $\mathcal{F}_{\lookupmulti}$}

\end{figure}

\begin{theorem}
\label{theorem:lookupsingle}
Protocol  $\Pi_{\hawksingle}^\online$ in Algorithm~\ref{algorithm:lookup-single-online} securely realizes the functionality $\mathcal{F}_{\lookupsingle}$ against semi-honest adversaries.
\end{theorem}
\begin{proofsketch}
Consider an adversary $\mathcal{A}$ that corrupts either $P_0$ or $P_1$. Without loss of generality, let's focus on the case where $\mathcal{A}$ corrupts $P_0$. We construct a simulator $\mathcal{S}$ that simulates $\mathcal{A}$ in the ideal world. The simulator $\mathcal{S}$ runs $\mathcal{A}$. $\mathcal{S}$ generates a random value $r$ to mimic the honest server $P_1$'s message $\share{x}{1} + H(k_1 || c) \pmod{n}$ and forwards it to $\mathcal{A}$. To simulate the honest server's output, $\mathcal{S}$ uses $\share{y}{1}$ received from $\mathcal{F}_{\lookupsingle}$.

The indistinguishability between the real and ideal worlds stems from the fact that the sole message exchanged, $\share{x}{1} + H(k_1 || c) \pmod{n}$, is computationally indistinguishable from a random value $r$, owing to the one-time pseudorandom pad $H(k_1 || c)$.
\end{proofsketch}

\begin{theorem}
\label{theorem:shareconvert}
Protocol $\Pi_{\mathsf{SC}} ^ {\online}$ in Algorithm~\ref{algorithm:share-convert} securely realizes the functionality $\mathcal{F}_{\mathsf{SC}}$ against semi-honest adversaries.
\end{theorem} 
\vspace{-3mm}

\begin{proofsketch}
Given the symmetry of our protocol with respect to both servers, we focus on the scenario where an adversary corrupts party $P_0$. To simulate the honest server $P_1$ in the ideal world, the simulator $\mathcal{S}$ generates a random value $s$ representing the message $\share{x}{1}^n + \share{r}{1}^n \pmod{n}$ and sends it to $\mathcal{A}$. $\mathcal{S}$ employs $\share{x}{1}^N$ obtained from $\mathcal{F}_{\mathsf{SC}}$  to simulate the honest server's output.

The indistinguishability between the real and ideal worlds stems from the fact that the sole message exchanged is $\share{x}{1}^n + \share{r}{1}^n \pmod{n}$, with $\share{r}{1}^n$ being a random share of $r$. Consequently, $\share{x}{1}^n + \share{r}{1}^n \pmod{n}$ is computationally indistinguishable from a random $s$ of equivalent length.
\end{proofsketch}

\subsection{Leakage Analysis}
\label{sec:leakageanalysis}

We begin by discussing the leakage function $\mathcal{L}$ for our $\hawkmulti$ protocol. Intuitively, our $\hawkmulti$ protocol leaks a perturbed profile of access patterns to a lookup table $L^{f,c}$, where $c$ is the table index. Note that the protocol does not reveal the input $x$ to any party. Instead, it only reveals a distorted frequency distribution of accesses to a table $L^{f,c}$ without associating access to a specific input $x$ due to the pseudorandom nature of the key $\kappa$. For instance, if we access a lookup table $\text{r}_{\text{Multi}} = 10$ times, each with $\epsilon$-$d_{\mathcal{X}}$-privacy, before exhausting our total privacy budget $\epsilon_T$, the $\hawkmulti$ protocol merely discloses if the ten lookups were distinct or if there were repeated lookups for a particular key $\kappa$. Furthermore, adding noise $\gamma$ to $x$  before each lookup ensures that parties remain unaware if a key repetition was genuine or noise-induced. We employ the geometric mechanism in Definition~\ref{def:geometric_mechanism} to sample the noise $\gamma$.

\begin{definition}
\label{def:leakagefndef}
The leakage function $\mathcal{L}$ is $\kappa$, where $\kappa$ is a pseudorandom key for a lookup table $L^{f,c}$ and
 $\kappa = k_{0}k_{1}( \hat{x}G + s_{0}^{c}G+ s_{1}^{c}G)$, where $k_{i}$ and $s_{i}^{c}$ are secret random keys of a party $P_i$, $G$ is the generator of an elliptic curve $E$, and $\hat{x} = x + \gamma$, with $\gamma$ being random noise as per Definition~\ref{def:geometric_mechanism}.
\end{definition}

\begin{theorem}
\label{theorem:dpleakage}
The leakage function $\mathcal{L}$ is local $\epsilon$-$d_{\mathcal{X}}$-private as per Definition~\ref{def:local_dx_privacy}.
\end{theorem} 
\vspace{-3mm}

\noindent  The preservation of local $\epsilon$-$d_{\mathcal{X}}$-privacy by $\mathcal{L}$ is a consequence of our utilization of the geometric mechanism, as defined in Definition~\ref{def:geometric_mechanism}, to sample noise $\gamma$. We prove this mechanism to preserve $\epsilon$-$d_{\mathcal{X}}$-privacy in Appendix~\ref{app:proofs}. Given the post-processing resilience of differential privacy~\cite{Fernandes2022,diff_priv_book}, $\mathcal{L}$ also maintains $\epsilon$-$d_{\mathcal{X}}$-privacy.

\begin{theorem}
\label{theorem:lookupmulti}
Protocol $\Pi_{\hawkmulti}^\online$ in Algorithm~\ref{algorithm:lookup-multi-online} securely realizes the functionality $\mathcal{F}_{\lookupmulti}$ with $\mathcal{L}$ leakage in ($\mathcal{F}_{\mathsf{SC}}$)-hybrid model as per Definition~\ref{def:hybridmodel}.
\end{theorem} 
\vspace{-3mm}

\begin{proofsketch}
Consider an adversary $\mathcal{A}$ that corrupts either server $P_0$ or $P_1$. Given the protocol's symmetry concerning both servers, we focus on the scenario where $\mathcal{A}$ corrupts $P_0$. We describe a simulator $\mathcal{S}$ to simulate $\mathcal{A}$ in the ideal world. The simulator $\mathcal{S}$ runs $\mathcal{A}$. $\mathcal{S}$ employs random $s$ to simulate the honest server $P_1$'s output of $\mathcal{F}_{\mathsf{SC}}$. $\mathcal{S}$ also generates random points $R$ and $S$ on the elliptic curve $E$ to simulate $P_1$'s messages $k_{1}( \share{\hat{x}}{1}^{N}G + s^{c}_{1}G )$ and $k_{1}k_{0}( \share{\hat{x}}{0}^{N}G + s^{c}_{0}G )$ and sends it to $\mathcal{A}$. $\mathcal{S}$ also sends leakage function $\mathcal{L}$ in Definition $\ref{def:leakagefndef}$ to $\mathcal{A}$. $\mathcal{S}$ uses $\share{y}{1}$ received from $\mathcal{F}_{\lookupmulti}$ to simulate the honest server's output.

The adversary $\mathcal{A}$'s view in the ideal and real worlds is indistinguishable because $\share{x}{1}^N$ is computationally indistinguishable from a random value $s$. Moreover, according to the \textit{elliptic curve discrete log assumption} $k_{1}( \share{\hat{x}}{1}^{N}G + s^{c}_{1}G )$ and $k_{1}k_{0}( \share{\hat{x}}{0}^{N}G + s^{c}_{0}G )$ are indistinguishable from random points $R$ and $S$ on $E$ as $\share{\hat{x}}{0}^{N}$, $\share{\hat{x}}{1}^{N}$, $k_0$, $k_1$, $s_0^c$, and $s_1^c$ are uniform random in \( \mathbb{Z}_N \). The only leakage in the real world and the ideal world is $\mathcal{L}$ as defined in Definition $\ref{def:leakagefndef}$ which is proved to preserve $\epsilon$-$d_{\mathcal{X}}$-privacy. 
\end{proofsketch}

%% file: rounds_and_communication_merged.tex
\begin{table}
\resizebox{85mm}{!}{%
\begin{tabular}{|l|c|c|l|l|}
\hline
\multirow{2}{*}{\textbf{Protocol}} & \multicolumn{2}{c|}{\textbf{Rounds}}                                       & \multicolumn{2}{c|}{\textbf{Communication}}                                \\ \cline{2-5} 
                                   & \multicolumn{1}{l|}{$\textsc{Hawk}_{\text{Single}}$} & \multicolumn{1}{l|}{$\textsc{Hawk}_{\text{Multi}}$} & \multicolumn{1}{l|}{$\textsc{Hawk}_{\text{Single}}$} & \multicolumn{1}{l|}{$\textsc{Hawk}_{\text{Multi}}$} \\ \hline                                      
 $\Pi _\lookup$        &        1         &  3 &      $l$  &  $l+ 2l'$            \\ \hline
%\multirow{2}{*}{$\Pi_{\relu(P_0, P_1)}$}         &        \multirow{2}{*}{  1  }     & \multirow{2}{*}{1} &      \multirow{2}{*}{$l$}  & \multirow{2}{*}{ $l$ }                     \\ 
% &&&& \\ 

%(after $\drelu$) &&&& \\ \hline

$\Pi_{\drelu}$                      &         1       &    3 &      $l$     &          $l+ 2l'$           \\ \hline  

$\Pi_{\relu}$  (after $\drelu$)       &         1       & 1 &      $l$  & $l$                      \\ \hline

$\Pi_{\sigmoid}$                        &       1         &   3 &    $l$    &  $l+ 2l'$                  \\ \hline
$\Pi_{\softmax}$                       &       3         & 7 &         $l$(1 + 2$d_l$)   &         $l(1+2d_{l})+2l'(1+d_{l})$             \\ \hline 
    
\end{tabular}
}
\caption{Total rounds and communication complexity of $\textsc{Hawk}_{\text{Single}}$ and $\textsc{Hawk}_{\text{Multi}}$ protocols. (\textbf{$l$} = 64 bits, \textbf{$l^{'}$} = 256 bits)}
\label{table:rounds_and_complexity}
\end{table}

%% file: privacy_preserving_ml.tex
\section{End-to-end Training Protocols}
\label{sec:privacy_preserving_ml}

\subsection{Privacy-Preserving Logistic Regression}
\label{sec:logistic_reg}

Logistic regression necessitates the computation of the $\sigmoid$ function. While previous studies have approximated $\sigmoid$ using polynomials~\cite{logistic_reg_polynomial}, achieving satisfactory accuracy demands high-degree polynomials, leading to inefficiencies. SecureML uses a piece-wise function and Yao’s garbled circuits for Sigmoid approximation, but the overhead and accuracy loss from this method remain a concern. We introduce the $\Pi_{\sigmoid}$ protocol to directly compute $\sigmoid$, enabling efficient training of logistic regression models without compromising plaintext training accuracy. Algorithm~\ref{algorithm:logistic_regression} in Appendix~\ref{app:lr_protocol} presents our complete PPLR protocol. After the dot product computation of input examples with coefficients during the forward pass, parties employ $\Pi_{\sigmoid}$ for $\sigmoid$ computation, subsequently proceeding to backward propagation.

\subsection{Privacy-Preserving Neural Network}

Training neural networks securely introduces complexity and inefficiency primarily during the computation of non-linear activations, such as $\relu$, $\drelu$, and $\softmax$. SecureML employs Yao's garbled circuits for $\relu$ and $\drelu$ computations. They also suggest an approximation for $\softmax$ as $f(z_i) = \frac{ReLU(z_i)}{\sum_{j=1}^{d_l}ReLU(z_j)}$. However, this function demands a division garbled circuit, leading to inefficiencies, and the approximation compromises accuracy, as highlighted in Section~\ref{subsec:neural_network_training}.

In Section~\ref{subsec:Act-Func}, we introduce efficient protocols for computing $\relu$, $\drelu$, and $\softmax$. We can adeptly train diverse neural networks by integrating the methods from Section~\ref{sec:logistic_reg} and the protocols in Section~\ref{subsec:Act-Func}. For instance, SecureML's 3-layer neural network, comprising two fully connected layers with $\relu$ activation and an output layer with $\softmax$ activation, can be implemented in the forward pass using $\Pi_{\secmul}$, $\Pi_{\relu}$, and $\Pi_{\softmax}$. Back-propagation, on the other hand, requires $\Pi_{\secmul}$ and $\Pi_{\drelu}$. Although our framework is readily extendable to CNNs, we focus on DNNs in our evaluation and defer CNNs to future work.

%% file: evaluation.tex
\section{Evaluation}
\label{sec:evaluation}

\subsection{Implementation}
We implement our system in C++ with $\sim13,570$ lines of code. We use the Eigen library~\cite{eigenweb} for matrix operations, libsecp256k1~\cite{secp256k1} for EC implementation, and Crypto++~\cite{crptopp} for cryptographic operations.

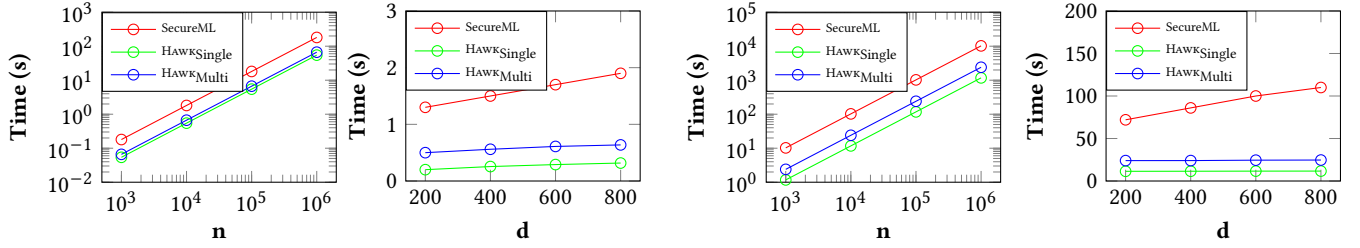
\begin{figure*}
    \centering
    
    % 1st Graph
    \begin{subfigure}{0.24\textwidth}
        \begin{tikzpicture}
            \begin{axis}[
                xlabel=\textbf{n},
                ylabel=\textbf{Time (s)},
                xmode=log,
                ymode=log,
                log ticks with fixed point,
                xtick={1000,10000,100000,1000000},
                xticklabels={\(10^3\),\(10^4\),\(10^5\),\(10^6\)},
                ytick={0.01,0.1,1,10,100, 1000},
                yticklabels={\(10^{-2}\),\(10^{-1}\),\(10^0\),\(10^1\),\(10^2\),\(10^3\)},
                ymin=0.01, ymax=1000,
                %legend pos=north west,
                legend style={at={(0.0,1)}, anchor=north west, font=\fontsize{5}{5}\selectfont, text width=0.9cm},
                %legend image code/.code={},
                ]
                \addplot[red,mark=o] coordinates {(10^3, 0.18) (10^4, 1.809) (10^5, 18.09) (10^6, 180.95)};
                \addplot[green,mark=o] coordinates {(10^3,0.054) (10^4, 0.546) (10^5, 5.46) (10^6, 54.68)};
                \addplot[blue,mark=o] coordinates {(10^3,0.066) (10^4,0.667) (10^5, 6.67) (10^6, 66.7)};
                \legend{SecureML,$\textsc{Hawk}_{\text{Single}}$,$\textsc{Hawk}_{\text{Multi}}$}
            \end{axis}
        \end{tikzpicture}
        %\caption{LAN setting (d=784)}
    \end{subfigure}\hfill
    % 2nd Graph
    \begin{subfigure}{0.24\textwidth}
        \begin{tikzpicture}
            \begin{axis}[
                xlabel=\textbf{d},
                ylabel=\textbf{Time (s)},
                ytick={0,1,2,3},
                ymin=0, ymax=3,
                %legend pos=north west,
                legend style={at={(0.0,1)}, anchor=north west,font=\fontsize{5}{5}\selectfont, text width=0.9cm},
                %legend image code/.code={},
                ]
                \addplot[red,mark=o] coordinates {(200, 1.3) (400, 1.5) (600, 1.7) (800, 1.9)};
                \addplot[green,mark=o] coordinates {(200, 0.2) (400, 0.255) (600, 0.29) (800, 0.317)};
                \addplot[blue,mark=o] coordinates {(200, 0.5) (400, 0.56) (600, 0.61) (800, 0.637)};
                \legend{SecureML,$\textsc{Hawk}_{\text{Single}}$,$\textsc{Hawk}_{\text{Multi}}$}
            \end{axis}
        \end{tikzpicture}
        %\caption{LAN setting (n=10,000)}
    \end{subfigure}\hfill
    % 3rd Graph
    \begin{subfigure}{0.24\textwidth}
        \begin{tikzpicture}
            \begin{axis}[
                xlabel=\textbf{n},
                ylabel=\textbf{Time (s)},
                xmode=log,
                ymode=log,
                log ticks with fixed point,
                xtick={1000,10000,100000,1000000},
                xticklabels={\(10^3\),\(10^4\),\(10^5\),\(10^6\)},
                ytick={1,10,100,1000,10000, 100000},
                yticklabels={\(10^0\),\(10^1\),\(10^2\),\(10^3\),\(10^4\),\(10^5\)},
                ymin=1, ymax=100000,
                %legend pos=north west,
                legend style={at={(0.0,1)}, anchor=north west,font=\fontsize{5}{5}\selectfont, text width=0.9cm},
                %legend image code/.code={},
                ]
                \addplot[red,mark=o] coordinates {(10^3, 10.229) (10^4, 102.29) (10^5, 1022.93) (10^6, 10229.36)};
                \addplot[green,mark=o] coordinates {(10^3,1.17) (10^4, 11.71) (10^5, 117.18) (10^6, 1171.87)};
                \addplot[blue,mark=o] coordinates {(10^3,2.39) (10^4, 23.99) (10^5, 239.98) (10^6, 2399.89)};
                \legend{SecureML,$\textsc{Hawk}_{\text{Single}}$,$\textsc{Hawk}_{\text{Multi}}$}
            \end{axis}
        \end{tikzpicture}
        %\caption{WAN setting (d=784)}
    \end{subfigure}\hfill
    % 4th Graph
    \begin{subfigure}{0.24\textwidth}
        \begin{tikzpicture}
            \begin{axis}[
                xlabel=\textbf{d},
                ylabel=\textbf{Time (s)},
                ytick={0,50,100,150,200},
                ymin=0, ymax=200,
                %legend pos=north west,
                legend style={at={(0.0,1)}, anchor=north west,font=\fontsize{5}{5}\selectfont, text width=0.9cm},
                %legend image code/.code={},
                ]
                \addplot[red,mark=o] coordinates {(200, 72) (400, 86) (600, 100) (800, 110)};
                \addplot[green,mark=o] coordinates {(200, 11.42) (400, 11.5) (600, 11.63) (800, 11.68)};
                \addplot[blue,mark=o] coordinates {(200, 24) (400, 24.03) (600, 24.51) (800, 24.67)};
               \legend{SecureML,$\textsc{Hawk}_{\text{Single}}$,$\textsc{Hawk}_{\text{Multi}}$}
            \end{axis}
        \end{tikzpicture}
        %\caption{WAN setting (n=10,000)}
    \end{subfigure}

    \caption{a) LAN setting (d=784) \hfill b) LAN setting (n=10,000) \hfill c) WAN setting (d=784) \hfill d) WAN setting (n=10,000) \\
    Comparison of our PPLR protocols with SecureML. Here $\textbf{n}$ denotes number of data records, and $\textbf{d}$ denotes the data dimension.}
    \label{fig:LR_scalability_multi_client}
\end{figure*}

%\subsection{Number Encoding}
We adopt a field size of $2^{64}$ and represent numbers using C++'s native unsigned long integer datatype. Following SecureML's fixed-point representation~\cite{secureml}, we allocate 13 bits to the fractional component. For lookup tables, numbers are represented with 16 bits divided between integer and fractional parts. The lookup tables for $\sigmoid$ allocate 3 bits for the integer and 13 for the fractional part. For $\drelu$ tables, based on the alternate representation from Section~\ref{subsec:single_use_lookup}, we use 4 integer bits and 1 fractional bit. For the $\softmax$ tables, the exponential function is represented with 6 integer and 10 fractional bits, while the inverse function uses 14 integer and 2 fractional bits, ensuring the necessary high dynamic range. In our experiments involving $\hawkmulti$ protocol, $N$ is set to match the order of the SECP256K1 curve.

\subsection{Experimental Setup}

Our experiments utilize two ``Amazon EC2 c4.8x large'' instances, aligning with the setup in previous work~\cite{secureml}. We assess performance in both LAN and WAN settings:

\noindent \textbf{LAN setting:} Both "Amazon EC2 c4.8x large" instances are hosted within the same region, achieving an average bandwidth of 620 MB/s and a network delay of 0.25ms.

\noindent \textbf{WAN setting:} The instances are hosted in separate regions: US East and US West. This configuration yields an average bandwidth of 32 MB/s and a network delay of 48 ms.

%\subsection{Datasets}

We use these datasets to evaluate our protocols: MNIST~\cite{mnist}, Gisette~\cite{gisette, arcene_paper}, Arcene~\cite{arcene, arcene_paper}, Fashion MNIST~\cite{fashion-mnist}, Iris~\cite{iris-dataset-1}, and the Adult dataset~\cite{adult}. We describe each dataset in Appendix~\ref{app:datasets}.

%\subsection{Experiments Summary}
Each data point represents the average of 10 runs. To assess scalability, we create synthetic datasets from MNIST by altering record count and data dimensions.
%We include execution times for our protocols' offline and online phases and present secure inference benchmarks in Section~\ref{subsec:secure_inference}.
For our $\textsc{Hawk}_{\text{Multi}}$ protocol, we experiment by adjusting the total privacy budget, $\epsilon_{T}$, of the lookup tables, where $\epsilon_{T} = \epsilon \times \text{r}_{\text{Multi}}$, with $\epsilon$ as the per-access privacy budget and $\text{r}_{\text{Multi}}$ denoting each lookup table's reuse frequency.

\subsection{Privacy Preserving Logistic Regression}
\label{subsec:logistic_regression_training}  	
This section evaluates our PPLR protocol's performance and scalability, compares it with SecureML~\cite{secureml}, and assesses the accuracy against plaintext training.
\input{accuracy_comparison_LR}

\begin{figure}
    \centering
    
    % 1st Graph
    \begin{subfigure}{0.27\textwidth}
        \begin{tikzpicture}
            \begin{axis}[
                xlabel=\textbf{Epochs},
                ylabel=\textbf{Accuracy (\%)},
                xtick={0,2,4,6,8, 10},
                xticklabels={0,2,4,6,8,10},
                ytick={97.5, 98.00, 98.50, 99.00, 99.50},
                yticklabels={97.5, 98, 98.5, 99, 99.5},
                ymin=97.50, ymax=99.50,
                legend pos=south east,
                ]
                \addplot[green,mark=o] coordinates {(1,98.4033) (2,98.8133) (3,98.9333) (4,99.02) (5,99.0767) (6,99.1) (7,99.1467) (8,99.1733) (9,99.2) (10,99.2133)};
                \addplot[blue,mark=o] coordinates {(1,98.3967) (2,98.82) (3,98.9433) (4,99.0167) (5,99.0767) (6,99.0967) (7,99.13) (8,99.1733) (9,99.1933) (10,99.2133)};
                \addplot[brown,mark=o] coordinates {(1,98.1933) (2,98.66) (3,98.85) (4,98.9333) (5,98.9767) (6,99.03) (7,99.0767) (8,99.1033) (9,99.1067) (10,99.1567)};
                \addplot[red,mark=o] coordinates {(1,97.9033) (2,98.4333) (3,98.7133) (4,98.83) (5,98.91) (6,98.9167) (7,98.9533) (8,98.9667) (9,99.0) (10,99.0167)};
                \legend{$\epsilon_{T} = \infty$,$\epsilon_{T} = 0.01$,$\epsilon_{T} = 0.001$,$\epsilon_{T} = 0.0006$}
            \end{axis}
        \end{tikzpicture}
        %\caption{MNIST dataset}
    \end{subfigure}\hfill
    \begin{subfigure}{0.27\textwidth}
        \begin{tikzpicture}
            \begin{axis}[
                xlabel=\textbf{Epochs},
                ylabel=\textbf{Accuracy (\%)},
                xtick={0,2,4,6,8, 10},
                xticklabels={0,2,4,6,8,10},
                ytick={91, 92, 93, 94, 95, 96, 97},
                yticklabels={91, 92, 93, 94, 95, 96, 97},
                ymin=91, ymax=97,
                legend pos=south east,
                ]
                \addplot[green,mark=o] coordinates {(1,92.4333) (2,94.7667) (3,95.4) (4,95.9) (5,96.1333) (6,96.2) (7,96.2667) (8,96.6333) (9,96.8) (10,96.9333)};
                \addplot[blue,mark=o] coordinates {(1,92.4667) (2,94.8333) (3,95.4) (4,95.9667) (5,96.2) (6,96.2) (7,96.2) (8,96.4) (9,96.6667) (10,96.9333)};
                \addplot[brown,mark=o] coordinates {(1,92.0) (2,93.7667) (3,94.8667) (4,95.1667) (5,95.7) (6,96.0333) (7,96.1667) (8,96.2) (9,96.3333) (10,96.4)};
                \addplot[red,mark=o] coordinates {(1,91.7) (2,93.8) (3,94.7) (4,95.2) (5,95.3) (6,95.6) (7,95.7) (8,95.7) (9,95.7) (10,95.4)};
                
                \legend{$\epsilon_{T} = \infty$,$\epsilon_{T} = 0.01$,$\epsilon_{T} = 0.001$,$\epsilon_{T} = 0.0006$}
            \end{axis}
        \end{tikzpicture}
       % \caption{Gissette dataset}
    \end{subfigure}

    \caption{a) MNIST dataset \hfill b) Gissette dataset \\
    \boldmath Training trajectory for PPLR across various $\epsilon_{T}$ values}
    \label{fig:epsilon_vs_accuracy_lr}
\end{figure}
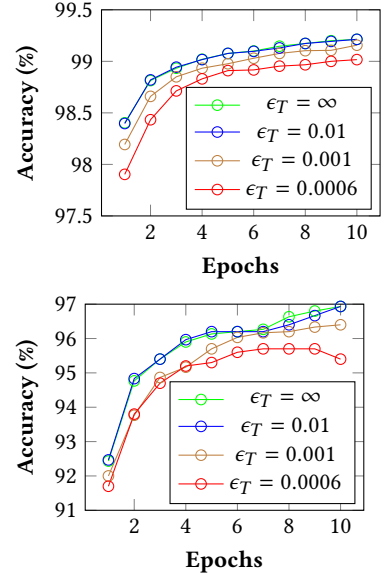

\input{training_time_comparison_logistic_multi_client}

\noindent \textbf{Accuracy Performance.}
Table \ref{table:accuracy_comparison_LR} indicates our $\textsc{Hawk}_{\text{Single}}$ protocol matches plaintext training accuracy across various datasets. For $\textsc{Hawk}_{\text{Multi}}$, there is no accuracy loss at $\epsilon_{T}=0.01$ and only a minor decline at $\epsilon_{T}=0.0005$, suggesting that $\textsc{Hawk}_{\text{Multi}}$ offers robust privacy without significant accuracy compromises. Figure~\ref{fig:epsilon_vs_accuracy_lr} illustrates the training trajectory on MNIST and Gissette datasets with $\textsc{Hawk}_{\text{Multi}}$, highlighting the influence of varying $\epsilon_{T}$. Comparing with SecureML~\cite{secureml}, in the task of classifying MNIST digits as zero vs. non-zero, our protocols achieve 99.21\% accuracy, closely followed by a Tensorflow model at 99.19\%. In contrast, SecureML reaches 98.62\%. Notably, SecureML's accuracy suffers more from non-standard activations, especially with neural networks, with a drop of $\sim3\%$ for the MNIST dataset, as discussed in Section \ref{subsec:neural_network_training}.

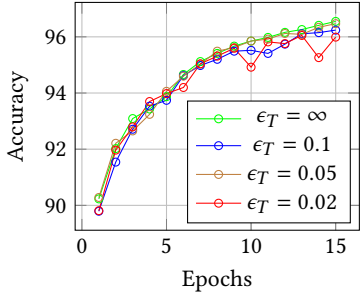
\begin{figure}
        \begin{tikzpicture}
            \begin{axis}[scale=0.21,
                xlabel={Epochs},
                ylabel={Accuracy},
                grid=major,
                legend pos = south east,
                %legend image code/.code={},
                ]
            \addplot[green,mark=o, mark size = 1.5 pt] coordinates {
            (1, 90.23) (2, 92.02) (3, 93.09) (4, 93.43) (5, 93.87) (6, 94.66) (7, 95.12) (8, 95.41) (9, 95.67) (10, 95.85) (11, 95.98) (12, 96.16) (13, 96.26) (14, 96.41) (15, 96.55)
            };
            \addplot[blue,mark=o, mark size = 1.5 pt] coordinates {
            (1, 89.79) (2, 91.54) (3, 92.71) (4, 93.53) (5, 93.74) (6, 94.62) (7, 94.98) (8, 95.20) (9, 95.49) (10, 95.52) (11, 95.41) (12, 95.74) (13, 96.11) (14, 96.15) (15, 96.24)
            };
            \addplot[brown,mark=o, mark size = 1.5 pt] coordinates {
            (1, 90.28) (2, 92.21) (3, 92.65) (4, 93.24) (5, 94.06) (6, 94.60) (7, 95.03) (8, 95.49) (9, 95.62) (10, 95.85) (11, 95.92) (12, 96.12) (13, 96.09) (14, 96.34) (15, 96.47)
            };
            \addplot[red,mark=o, mark size = 1.5 pt] coordinates {
            (1, 89.81) (2, 91.9667) (3, 92.7933) (4, 93.6967) (5, 93.98) (6, 94.1967) (7, 95.04) (8, 95.3133) (9, 95.5933) (10, 94.92) (11, 95.82) (12, 95.76) (13, 96.05) (14, 95.26) (15, 95.99)
            };
    \legend{$\epsilon_{T} = \infty$,$\epsilon_{T} = 0.1$,$\epsilon_{T} = 0.05$,$\epsilon_{T} =  0.02$}
            \end{axis}
        \end{tikzpicture}
        \caption{Training trajectory for PPNN across various \boldmath$\epsilon_{T}$ values for MNIST.}
        \label{fig:epsilon_vs_NN_train}
    \end{figure}

    \begin{figure}
        \begin{tikzpicture}
            \begin{axis}[scale=0.21,
                xlabel={Epochs},
                ylabel={Accuracy},
                grid=major,
                legend pos=north east,
                cycle list={
                    {green,mark=o},
                    %{green,mark=square},
                    {green,mark=triangle},
                    {blue,mark=o},
                    %{blue,mark=square},
                    {blue,mark=triangle},
                    {red,mark=o},
                    %{red,mark=square},
                    {red,mark=triangle},
                },
                legend style={fill opacity=0.3, draw=none, text opacity=1, at={(0.21,0.78)}, anchor=north west, font=\fontsize{6}{4}\selectfont, text width=2.3cm, row sep=-3pt},
                %legend image code/.code={},
                ]
\addplot+[each nth point={4}, filter discard/.code={\pgfmathparse{int(mod(\coordindex,4))}\ifnum\pgfmathresult=0\else\pgfmathfloatset\pgfmathresult{nan}\fi}] coordinates {
(1,10) (2,90.56) (3,92.27) (4,93.23) (5,93.89) (6,94.5) (7,94.93) (8,95.32) (9,95.62) (10,95.8) (11,95.95) (12,96.1) (13,96.24) (14,96.39) (15,96.47) (16,96.64) (17,96.39) (18,96.47) (19,96.64) (20,96.39) (21,96.47) (22,96.64) (23,96.39) (24,96.47) (25,96.64) (26,96.39) (27,96.39) (28,96.47) (29,96.64) (30,96.39)
};
%\addplot+[each nth point={4}, filter discard/.code={\pgfmathparse{int(mod(\coordindex,4))}\ifnum\pgfmathresult=0\else\pgfmathfloatset\pgfmathresult{nan}\fi}] coordinates {(1,10) (2,83.06) (3,87.95) (4,89.58) (5,90.45) (6,91.05) (7,91.66) (8,92.07) (9,92.43) (10,92.7) (11,92.87) (12,93.12) (13,93.34) (14,93.45) (15,93.61) (16,93.8) (17,93.34) (18,93.45) (19,93.61) (20,93.8) (21,93.34) (22,93.45) (23,93.61) (24,93.8) (25,93.34) (26,93.45) (27,93.61) (28,93.8) (29,93.34) (30,93.45)};
\addplot+[each nth point={4}, filter discard/.code={\pgfmathparse{int(mod(\coordindex,4))}\ifnum\pgfmathresult=0\else\pgfmathfloatset\pgfmathresult{nan}\fi}] coordinates {
(1,10) (2,53.95) (3,72.22) (4,79.22) (5,82.95) (6,84.99) (7,86.56) (8,87.53) (9,88.24) (10,88.68) (11,89.23) (12,89.52) (13,89.79) (14,90.01) (15,90.28) (16,90.48) (17,90.6) (18,90.84) (19,91.02) (20,91.17) (21,91.34) (22,91.41) (23,91.42) (24,91.53) (25,91.61) (26,91.63) (27,91.42) (28,91.53) (29,91.61) (30,91.63)
};
\addplot+[each nth point={4}, filter discard/.code={\pgfmathparse{int(mod(\coordindex,4))}\ifnum\pgfmathresult=0\else\pgfmathfloatset\pgfmathresult{nan}\fi}] coordinates {
(1,10) (2,89.91) (3,91.94) (4,92.91) (5,93.57) (6,94.19) (7,94.6) (8,95.1) (9,95.33) (10,95.47) (11,95.72) (12,95.88) (13,96.1) (14,95.92) (15,96.17) (16,95.9) (17,96.1) (18,95.92) (19,96.17) (20,95.9) (21,96.1) (22,95.92) (23,96.17) (24,95.9) (25,96.1) (26,95.92) (27,96.17) (28,95.9) (29,96.1) (30,95.92)
};
%\addplot+[each nth point={4}, filter discard/.code={\pgfmathparse{int(mod(\coordindex,4))}\ifnum\pgfmathresult=0\else\pgfmathfloatset\pgfmathresult{nan}\fi}] coordinates {(1,10) (2,81.82) (3,88.02) (4,89.62) (5,90.38) (6,90.83) (7,91.14) (8,91.55) (9,91.99) (10,92.27) (11,92.54) (12,92.72) (13,92.9) (14,93.21) (15,93.3) (16,93.6) (17,93.21) (18,93.3) (19,93.6) (20,93.21) (21,93.3) (22,93.6) (23,93.21) (24,93.3) (25,93.6) (26,93.21) (27,93.21) (28,93.3) (29,93.6) (30,93.21)};
\addplot+[each nth point={4}, filter discard/.code={\pgfmathparse{int(mod(\coordindex,4))}\ifnum\pgfmathresult=0\else\pgfmathfloatset\pgfmathresult{nan}\fi}] coordinates {
(1,10) (2,60.7) (3,74.16) (4,80.02) (5,83.43) (6,85.75) (7,86.94) (8,87.83) (9,88.61) (10,89.03) (11,89.46) (12,89.73) (13,90.01) (14,90.13) (15,90.31) (16,90.45) (17,90.58) (18,90.69) (19,90.91) (20,91.02) (21,91.13) (22,91.25) (23,91.38) (24,91.51) (25,91.61) (26,91.71) (27,91.38) (28,91.51) (29,91.61) (30,91.71)
};
% Softmax approx data
\addplot+[each nth point={4}, filter discard/.code={\pgfmathparse{int(mod(\coordindex,4))}\ifnum\pgfmathresult=0\else\pgfmathfloatset\pgfmathresult{nan}\fi}] coordinates {
(1,10) (2,82.29) (3,11.9) (4,11.93) (5,11.9) (6,10.13) (7,10.13) (8,10.13) (9,10.13) (10,10.13) (11,10.13) (12,10.13) (13,10.13) (14,10.13) (15,10.13) (16,10.13) (17,10.13) (18,10.13) (19,10.13) (20,10.13) (21,10.13) (22,10.13) (23,10.13) (24,10.13) (25,10.13) (26,10.13) (27,10.13) (28,10.13) (29,10.13) (30,10.13)
};
%\addplot+[each nth point={4}, filter discard/.code={\pgfmathparse{int(mod(\coordindex,4))}\ifnum\pgfmathresult=0\else\pgfmathfloatset\pgfmathresult{nan}\fi}] coordinates {(1,10) (2,77.32) (3,85.01) (4,87.2) (5,87.71) (6,87.86) (7,87.67) (8,86.84) (9,9.8) (10,9.77) (11,15.64) (12,19.23) (13,10.06) (14,13.35) (15,8.92) (16,10.1) (17,10.1) (18,10.1) (19,10.1) (20,10.1) (21,10.1) (22,10.1) (23,10.1) (24,10.1) (25,10.1) (26,10.1) (27,10.1) (28,10.1) (29,10.1) (30,10.1)};
\addplot+[each nth point={4}, filter discard/.code={\pgfmathparse{int(mod(\coordindex,4))}\ifnum\pgfmathresult=0\else\pgfmathfloatset\pgfmathresult{nan}\fi}] coordinates {
(1,10) (2,57.16) (3,72.69) (4,78.79) (5,81.6) (6,83.49) (7,84.72) (8,85.59) (9,85.94) (10,86.28) (11,86.5) (12,86.56) (13,86.68) (14,86.77) (15,86.81) (16,86.78) (17,86.68) (18,86.46) (19,86.23) (20,85.87) (21,85.48) (22,73.29) (23,10.31) (24,9.8) (25,9.8) (26,10.35) (27,10.31) (28,9.8) (29,9.8) (30,10.35)
};
                \legend{Tensorflow ($\alpha = 2^{-5}$),
    %Tensorflow ($\alpha = 2^{-7}$),
    Tensorflow ($\alpha = 2^{-9}$),
    $\textsc{Hawk}_{\text{Multi}}$ ($\alpha = 2^{-5}$),
    %$\textsc{Hawk}_{\text{Multi}}$ ($\alpha = 2^{-7}$),
    $\textsc{Hawk}_{\text{Multi}}$ ($\alpha = 2^{-9}$),
    Smax approx ($\alpha = 2^{-5}$),
    %Softmax approx ($\alpha = 2^{-7}$),
    Smax approx ($\alpha = 2^{-9}$)}
            \end{axis}
        \end{tikzpicture}
    \caption{Accuracy comparison of PPNN training on MNIST. Here \boldmath$\alpha$ denotes learning rate and for $\textsc{Hawk}_{\text{Multi}}$, $\epsilon_{T}=0.1$. The Softmax approximation used by past works causes accuracy drop.}
    \label{fig:convergence_comparison_NN}
\end{figure}
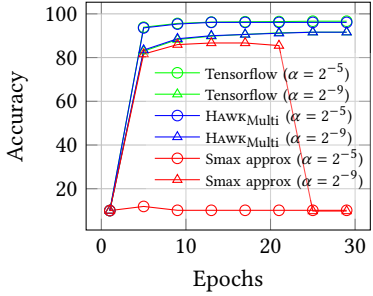

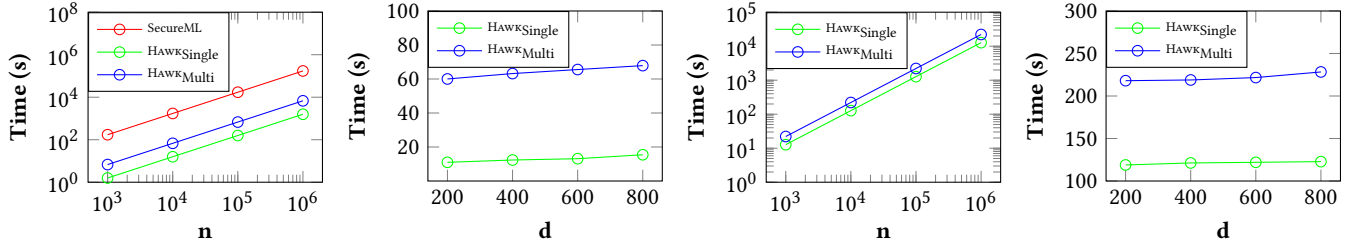
\begin{figure*}
    \centering
    
    % 1st Graph
    \begin{subfigure}{0.24\textwidth}
        \begin{tikzpicture}
            \begin{axis}[
                xlabel=\textbf{n},
                ylabel=\textbf{Time (s)},
                xmode=log,
                ymode=log,
                log ticks with fixed point,
                xtick={1000,10000,100000,1000000},
                xticklabels={\(10^3\),\(10^4\),\(10^5\),\(10^6\)},
                ytick={1,100,  10000,  1000000, 100000000},
                yticklabels={\(10^0\),\(10^2\), \(10^4\), \(10^6\), \(10^8\)},
                ymin=1, ymax=100000000,
                %legend pos=north west,
                legend style={at={(0.0,1)}, anchor=north west, font=\fontsize{5}{5}\selectfont, text width=0.9cm},
                %legend image code/.code={},
                ]
                \addplot[red,mark=o] coordinates {(10^3, 172) (10^4, 1720) (10^5, 17200) (10^6, 172000)};
                \addplot[green,mark=o] coordinates {(10^3,1.578) (10^4, 15.78) (10^5, 157.8) (10^6, 1578.12)};
                \addplot[blue,mark=o] coordinates {(10^3,6.772) (10^4, 67.72) (10^5, 677.24) (10^6, 6772.39)};
                \legend{SecureML,$\textsc{Hawk}_{\text{Single}}$,$\textsc{Hawk}_{\text{Multi}}$}
            \end{axis}
        \end{tikzpicture}
    \end{subfigure}\hfill
    % 2nd Graph
    \begin{subfigure}{0.24\textwidth}
        \begin{tikzpicture}
            \begin{axis}[
                xlabel=\textbf{d},
                ylabel=\textbf{Time (s)},
                ytick={20,40,60,80, 100},
                ymin=0, ymax=100,
                %legend pos=north west,
                legend style={at={(0.0,1)}, anchor=north west,font=\fontsize{5}{5}\selectfont, text width=0.9cm},
                %legend image code/.code={},
                ]
                \addplot[green,mark=o] coordinates {(200, 10.99) (400,  12.35) (600, 13.13) (800, 15.44)};
                \addplot[blue,mark=o] coordinates {(200, 60.02) (400,  63.2) (600,  65.56) (800, 67.87)};
                \legend{$\textsc{Hawk}_{\text{Single}}$,$\textsc{Hawk}_{\text{Multi}}$}
            \end{axis}
        \end{tikzpicture}
    \end{subfigure}\hfill
    % 3rd Graph
    \begin{subfigure}{0.24\textwidth}
        \begin{tikzpicture}
            \begin{axis}[
                xlabel=\textbf{n},
                ylabel=\textbf{Time (s)},
                xmode=log,
                ymode=log,
                log ticks with fixed point,
                xtick={1000,10000,100000,1000000},
                xticklabels={\(10^3\),\(10^4\),\(10^5\),\(10^6\)},
                ytick={1,10,100,1000,10000, 100000},
                yticklabels={\(10^0\),\(10^1\),\(10^2\),\(10^3\),\(10^4\),\(10^5\)},
                ymin=1, ymax=100000,
                %legend pos=north west,
                legend style={at={(0.0,1)}, anchor=north west,font=\fontsize{5}{5}\selectfont, text width=0.9cm},
                %legend image code/.code={},
                ]
                \addplot[green,mark=o] coordinates {(10^3,12.72) (10^4, 127.26) (10^5, 1272.65) (10^6, 12726.56)};
                \addplot[blue,mark=o] coordinates {(10^3,22.1) (10^4,   221.05) (10^5,  2210.5) (10^6, 22105.62)};
                \legend{$\textsc{Hawk}_{\text{Single}}$,$\textsc{Hawk}_{\text{Multi}}$}
            \end{axis}
        \end{tikzpicture}
    \end{subfigure}\hfill
    % 4th Graph
    \begin{subfigure}{0.24\textwidth}
        \begin{tikzpicture}
            \begin{axis}[
                xlabel=\textbf{d},
                ylabel=\textbf{Time (s)},
                ytick={100,150,200,250,300},
                ymin=100, ymax=300,
                %legend pos=north west,
                legend style={at={(0.0,1)}, anchor=north west,font=\fontsize{5}{5}\selectfont, text width=0.9cm},
                %legend image code/.code={},
                ]
                \addplot[green,mark=o] coordinates {(200, 118.88) (400, 121.25) (600,  121.99) (800,  122.81)};
                \addplot[blue,mark=o] coordinates {(200, 218.01) (400,  218.86) (600,  221.69) (800,  228.29)};
               \legend{$\textsc{Hawk}_{\text{Single}}$,$\textsc{Hawk}_{\text{Multi}}$}
            \end{axis}
        \end{tikzpicture}
    \end{subfigure}

    \caption{a) LAN setting (d=784) \hfill  b) LAN setting (n=10,000) \hfill c) WAN setting (d=784) \hfill  d) WAN setting (n=10,000) \\
    Comparison of our PPNN protocols with SecureML. Here, $\textbf{n}$ denotes number of data records, and $\textbf{d}$ denotes the data dimension.}
    \label{fig:NN_scalability_multi_client}
\end{figure*}

\noindent \textbf{Computation and Communication Performance.}
Table~\ref{table:training_time_comparison_logistic_multi_client} details our PPLR experiments in LAN and WAN environments. The online phase for $\textsc{Hawk}_{\text{Single}}$ requires 3.2s (LAN) and 70.2s (WAN) with a lookup table storage overhead of 36.62GB per party. In contrast, $\textsc{Hawk}_{\text{Multi}}$ has an online duration of 4s (LAN) and 144s (WAN) but reduces storage cost to 0.37GB for $\text{r}_{\text{Multi}} = 100$. This table underlines the balance between privacy, offline cost, and storage. While $\textsc{Hawk}_{\text{Single}}$ offers perfect security with a faster online phase, $\textsc{Hawk}_{\text{Multi}}$ demands fewer lookup tables, ensuring $d_{\mathcal{X}}$-privacy against access pattern leaks. Relative to SecureML, our methods are up to 3.4x and 8.7x faster in LAN and WAN, respectively.

For scalability experiments, we vary the dataset size $n$ and the data dimension $d$. Figure~\ref{fig:LR_scalability_multi_client} illustrates that PPLR training time scales linearly with $n$. As $d$ grows, our PPLR time increases sublinearly, while SecureML's rises linearly. In WAN, with 
$d$ ranging from 200 to 1000, SecureML's time increases from 72s to 124s. In contrast, our protocols remain steady at around 12s for $\textsc{Hawk}_{\text{Single}}$ and 24s for $\textsc{Hawk}_{\text{Multi}}$, as $d$ increases. In WAN, the overall time is largely influenced by communication time, which, for our protocols, stays nearly constant as $d$ increases.

\subsection{Privacy Preserving Neural Networks }
\label{subsec:neural_network_training}
To evaluate our protocols for training neural networks, we consider an architecture similar to prior works~\cite{secureml, securenn}. The neural network is fully connected, consisting of 2 hidden layers, each with 128 neurons and an output layer with 10 neurons. The hidden layers use $\relu$, and the output layer uses $\softmax$ activation.

\input{accuracy_comparison_NN}

\input{training_time_comparison_NN_multi_client}

\input{secure_inference_comparison_multi_client}

\noindent \textbf{Accuracy Performance.} The neural network trained on MNIST using our PPNN protocol can reach an accuracy of $96.6\%$ within 15 epochs, similar to a Tensorflow model. Neural networks trained on Iris~\cite{iris-dataset-1} and the Adult dataset~\cite{adult} using our $\textsc{Hawk}_{\text{Single}}$ protocol closely match plaintext training accuracies, as shown in Table~\ref{table:accuracy_comparison_NN}. However, with $\textsc{Hawk}_{\text{Multi}}$, accuracy slightly diminishes as $\epsilon_{T}$ decreases. For instance, at $\epsilon_{T}=0.02$, MNIST accuracy drops by $\sim 0.5\%$. Figure~\ref{fig:epsilon_vs_NN_train} depicts the MNIST training trajectory under $\textsc{Hawk}_{\text{Multi}}$, emphasizing the effect of different $\epsilon_{T}$ values. Additionally, Figure~\ref{fig:luts_vs_epsilon}  illustrates the correlation between the total number of lookup tables required and various privacy budget thresholds necessary to achieve specific accuracy levels. As expected $\textsc{Hawk}_{\text{Single}}$ protocol requires the highest number of lookup tables. For the $\textsc{Hawk}_{\text{Multi}}$ protocol, a decrease in $\epsilon_{T}$ in case of the Adult dataset necessitates an increase in the number of lookup tables to maintain the same level of accuracy.

Previous studies~\cite{secureml, securenn, mohassel2018aby3} report around $93.4\%$ accuracy on MNIST using the same setup. This discrepancy primarily stems from their use of a non-standard activation function for $\softmax$. Figure~\ref{fig:convergence_comparison_NN} contrasts our training protocol's accuracy with plaintext training. When we emulate the $\softmax$ approximation ($f(z_i) = \frac{ReLU(z_i)}{\sum_{j=1}^{d_l}ReLU(z_j)}$) from prior works~\cite{secureml, securenn, mohassel2018aby3} in a plaintext NN model, we notice a decline in accuracy and delayed convergence. Notably, our protocol surpasses $90\%$ accuracy on MNIST in under 1.06 epochs, while earlier works~\cite{secureml, securenn, mohassel2018aby3} need more epochs for comparable results. As shown in Figure~\ref{fig:convergence_comparison_NN}, our protocol benefits from a larger learning rate, leading to quicker convergence. In contrast, the $\softmax$ approximation can cause accuracy to plummet after a certain number of epochs. This decline, consistent across various learning rates, suggests that such approximations might be even more detrimental for more complex tasks, potentially causing substantial accuracy losses.

\noindent \textbf{Computation and Communication Performance.} Table~\ref{table:training_time_comparison_NN_multi_client} compares our PPNN protocols with SecureML~\cite{secureml}. While we also reference other works~\cite{quotient_paper, securenn, mohassel2018aby3, falcon_paper} that encompass both 2PC and 3PC protocols, a direct comparison is challenging due to differing settings, as highlighted in Section~\ref{sec:related-work}. In LAN, our PPNN protocol's online phase outperforms SecureML's client-aided approach by $26-144\times$. SecureML, in the WAN context, omits client-aided protocol training times, offering only metrics for the setting with server-generated random shares. This method increases offline costs but reduces online time through matrix multiplication speed gains. Our WAN-based PPNN protocol is $159-281\times$ faster than SecureML's server-aided variant. For the client-aided variant, factoring in a 2.45 slowdown from the LAN setting, our protocol's speed advantage over SecureML in WAN surges to $391-688\times$. While our PPNN protocol drastically reduces offline costs compared to SecureML, our $\textsc{Hawk}_{\text{Single}}$ adds a lookup table storage overhead of 407.5GB. In contrast, our $\textsc{Hawk}_{\text{Multi}}$ reduces this overhead to 4.1GB per party. Compared with Quotient~\cite{quotient_paper}, another 2PC protocol, our PPNN protocol's online phase is $79-436\times$ and $137-242\times$ faster in LAN and WAN, respectively. Our PPNN protocols even outperform several 3PC protocols, including SecureNN, ABY$^{3}$, and Falcon~\cite{mohassel2018aby3,securenn,falcon_paper}, boasting speeds up to $52\times$ in LAN and $37\times$ in WAN.

%To assess scalability, we varied dataset size $n$ and dimensions $d$ in both LAN and WAN settings.
Figure~\ref{fig:NN_scalability_multi_client} presents our scalability experiments. Our protocol's training times scale linearly with $n$ and sub-linearly with $d$. SecureML only provides results for the client-aided LAN setting.

\subsection{Secure Inference}
\label{subsec:secure_inference}

%\todo{add communication cost}

In this section, we evaluate the efficacy of our protocol for secure inference using the MNIST dataset~\cite{mnist}. Our $\textsc{Hawk}_{\text{Single}}$ protocol is better suited for secure inference than $\textsc{Hawk}_{\text{Multi}}$ due to its reduced online cost. The cost of secure inference using $\textsc{Hawk}_{\text{Single}}$ for both logistic regression and neural networks is shown in Table~\ref{table:secure_inference_multi_client}, alongside a comparison with SecureML~\cite{secureml}. The neural network architecture referenced is the same as discussed in Section~\ref{subsec:neural_network_training}.

SecureML~\cite{secureml} only presents the inference times for their server-aided protocol. For neural networks, they only report inference times using the \textit{square} activation function. While the \textit{square} activation is faster than $\relu$, it compromises slightly on accuracy, as highlighted in SecureML~\cite{secureml}. Our client-aided protocol has a higher online cost for dot product operation because matrix multiplication isn't feasible~\cite{secureml}. Moreover, we employ the $\relu$ activation function, ensuring the preservation of plaintext accuracy. In terms of secure inference, our $\textsc{Hawk}_{\text{Single}}$ protocol's online phase is up to $2.5\times$ faster than SecureML for both logistic regression and neural networks. Additionally, the offline phase of $\textsc{Hawk}_{\text{Single}}$ outperforms SecureML across most scenarios.

%% file: accuracy_comparison_LR.tex
\begin{table}[ht]
\resizebox{85mm}{!}{%
\begin{tabular}{|c|c|c|c|c|}
\hline
      \textbf{Framework}       & \textbf{MNIST} & \textbf{Gisette} & \textbf{Arcene} & \textbf{Fashion MNIST} \\ \hline
Tensorflow    &   99.19    &   98      &   86  &  95.97    \\ \hline
%\textbf{SecureML} &  98.62     &    97.7     &     86  & 95.95 \\ \hline
$\textsc{Hawk}_{\text{Single}}$ &  99.21     &    98.1     &     86  & 95.97 \\ \hline
%\textbf{This (MultiLookup) $\epsilon=1$ } &  99.21     &   98.1      &    86   & 95.97 \\ \hline
%\textbf{This (MultiLookup) $\epsilon=0.1$ } &  99.21     &    98.1     &   86    & 95.97  \\ \hline
$\textsc{Hawk}_{\text{Multi}}$ ($\epsilon_{T}=0.01)$  &   99.21    &  98.1       &    86   & 95.97 \\ \hline
$\textsc{Hawk}_{\text{Multi}}$ ($\epsilon_{T}=0.001)$  &    99.20   &  97.1        &   85   & 95.89  \\ \hline
$\textsc{Hawk}_{\text{Multi}}$ ($\epsilon_{T}=0.0005)$  &   98.80    &    95.2      &  85    &  95.51 \\ \hline
\end{tabular}
}
\caption{Accuracy comparison of our PPLR protocols with plaintext training $(\text{r}_{\text{Multi}} \textbf{= 10})$.}
\label{table:accuracy_comparison_LR}
\end{table}

%% file: training_time_comparison_logistic_multi_client.tex
\begin{table}[ht]
\centering
\resizebox{80mm}{!}
{%
\begin{tabular}{|c|c|c|c|c|}
\hline
\textbf{Framework} & \textbf{LAN (s)} & \textbf{WAN (s)} & \textbf{Offline (s)} & \textbf{Storage} \\  \hline
%SecureML  & 8.98 & 397.38  \\ \hline
SecureML  & 10.84  & 612.78 & - & - \\ \hline \hline

$\textsc{Hawk}_{\text{Single}}$ & 3.2 & 70.2 & 57.58 & 36.62 GB \\ \hline
$\textsc{Hawk}_{\text{Multi}}$ ($\epsilon_{T} = 0.01$) & 4 & 144 & 18.3  & 0.37 GB \\ \hline

\end{tabular}
}
\caption{Comparison of PPLR with SecureML on MNIST ($\textbf{|B|}$ = 128, $\text{r}_{\text{Multi}} \textbf{= 100}$). SecureML does not report the offline cost.}
\label{table:training_time_comparison_logistic_multi_client}
\end{table}

%% file: accuracy_comparison_NN.tex
\begin{table}[ht]
\centering
\resizebox{60mm}{!}{%
\begin{tabular}{|c|c|c|c|c|}
\hline
      \textbf{Framework}       & \textbf{MNIST} & \textbf{Iris} & \textbf{Adult} \\ \hline
Tensorflow    &  96.6   & 96.56 &  85.4 \\ \hline
$\textsc{Hawk}_{\text{Single}}$ &   96.6    &  95.56 &  85.4 \\ \hline
$\textsc{Hawk}_{\text{Multi}}$ $(\epsilon_{T}=0.1)$ &   96.34    &  96.56 & 85.4  \\ \hline
$\textsc{Hawk}_{\text{Multi}}$ $(\epsilon_{T}=0.05)$ &   96.09    &  96.56 & 85.2  \\ \hline
$\textsc{Hawk}_{\text{Multi}}$ $(\epsilon_{T}=0.02)$ &  95.99     &  96.56 & 85.2  \\ \hline
\end{tabular}
}
\caption{Accuracy comparison of our PPNN protocols with plaintext training$(\text{r}_{\text{Multi}} \text{= 10})$.}
\label{table:accuracy_comparison_NN}
\end{table}

%% file: training_time_comparison_NN_multi_client.tex
\definecolor{Gray}{gray}{0.9}

\begin{table*}
\centering
\resizebox{150mm}{!}{%
\begin{tabular}{|clccccc|}
\hline
\multirow{2}{*}{} & \multirow{2}{*}{\textbf{Framework}} & \multicolumn{2}{c}{\textbf{LAN}} & \multicolumn{3}{c|}{\textbf{WAN}} \\ \cline{3-7} 
& &  \textbf{Offline (hours)} & \textbf{Online (hours)}  & \textbf{Offline (hours)} & \textbf{Online (hours)} & \textbf{Lookup Table Storage}  \\ \hline
\rowcolor{Gray}
 & $\textsc{Hawk}_{\text{Single}}$  & 0.47  & 0.02   & 0.47 & 0.21  & 407.5 GB \\ \cline{2-7}
 \rowcolor{Gray}
\multirow{3}{*}{2PC} & $\textsc{Hawk}_{\text{Multi}}$ ($\epsilon_{T} = 1$)  &  0.35 &  0.11  & 0.35 & 0.37 &  4.1 GB \\ \cline{2-7}

& SecureML (server-aided)~\cite{secureml}  &  80.5 & 1.17  & 4277 & 59 & - \\ \cline{2-7}
& SecureML (client-aided)~\cite{secureml}  & 4.15 & 2.87  & - & - & - \\ \cline{2-7}

& Quotient~\cite{quotient_paper}  & - & 8.72  & - & 50.74 & - \\ \hline \hline

\multirow{3}{*}{3PC}  

& Falcon~\cite{falcon_paper}  & - & 0.17  & - & 3.76 & - \\ \cline{2-7}

& ABY$^3$~\cite{mohassel2018aby3} &  - & 0.75  & - & - & - \\ \cline{2-7}

& SecureNN~\cite{securenn} &  - & 1.03  & - & 7.83 & - \\ \hline

\end{tabular}
}

\caption{Comparison of privacy-preserving neural network training using MNIST dataset ($\textbf{|B|}$ = 128, $\text{r}_{\text{Multi}} = 100$).}
\label{table:training_time_comparison_NN_multi_client}

\end{table*}

%% file: secure_inference_comparison_multi_client.tex
\begin{table*}
\centering
\resizebox{115mm}{!}{%
\begin{tabular}{|c|cc|cc|cc|cc|} 
\hline
\multirow{3}{*}{\textbf{Framework}}                                                                  & \multicolumn{4}{c|}{\textbf{Logistic Regression}}                                                                                          & \multicolumn{4}{c|}{\textbf{Neural Networks}}                                                                                             \\ 
\cline{2-9}
                                                                    & \multicolumn{2}{c|}{\textbf{LAN (\textcolor{red}{ms})}} & \multicolumn{2}{c|}{\textbf{WAN (\textcolor{red}{ms})}}      & \multicolumn{2}{c|}{\textbf{LAN (\textcolor{red}{s})}} & \multicolumn{2}{c|}{\textbf{WAN (\textcolor{red}{s})}}       \\ 
\cline{2-5}\cline{6-9}
                                                                 & \textbf{Offline} & \textbf{Online}                      & \textbf{Offline} & \textbf{Online}                               & \textbf{Offline} & \textbf{Online}                     & \textbf{Offline} & \textbf{Online}                               \\ 
\hhline{|=========|}
SecureML \textcolor{red}{(server-aided)}    & 51               & 3.9                                  & 2010             & 429                                   & 13.8             & 0.20                                & 472              & 1.2                                   \\ 
\hhline{|=========|}
$\textsc{Hawk}_{\text{Single}}$                                     & 96               & 3.7                                  & 96               & 171                                                    & 1.43              & 0.08                                 & 1.43              & 0.82                                                \\ \hline
\end{tabular}}
\caption{Secure inference time comparison with SecureML on MNIST for 100 samples.}
\label{table:secure_inference_multi_client}
\end{table*}

%% file: conclusion.tex
\section{Conclusion}
\label{sec:conclusion}

In this work, we present novel lookup table protocols for PPML that achieve both accuracy and efficiency in activation function computation. We present $\textsc{Hawk}_{\text{Single}}$, offering leakage-free security, and $\textsc{Hawk}_{\text{Multi}}$, enabling table reuse at the cost of leakage which preserves $d_{\mathcal{X}}$-privacy. Our PPML protocols for logistic regression and neural networks demonstrate significant performance gains, reaching up to $688\times$ faster training for neural networks than SecureML~\cite{secureml} with plaintext accuracy. Beyond activation functions, our approach applies to various MPC applications, paving the way for wider adoption of PPML and secure collaborative data analysis.

\begin{acks}
This research received no specific grant from any funding agency in the public, commercial, or not-for-profit sectors.
\end{acks}

%% file: appendix.tex
\appendix

\section{Machine Learning}
\label{app:machine_learning}

\noindent \textbf{Stochastic Gradient Descent (SGD)} is a commonly used iterative algorithm to find the local minimum of a function. We use SGD to minimize the cross-entropy cost function and learn $\boldsymbol{w}$. The coefficient vector $\boldsymbol{w}$ is first initialized to 0 or random values. Then in each iteration, we randomly select a data point $(\boldsymbol{x_i},y_i)$ and update $\boldsymbol{w}$ as $w_j = w_j - \alpha \frac{\partial C_i(\boldsymbol{w})}{\partial w_j}$, where $\alpha$ is the learning rate that controls the step towards the minimum taken in each iteration. Substituting the cost function, we get the update rule: $w_j = w_j - \alpha(f(\boldsymbol{x_i.w}) - y_i)x_{ij}$.

\noindent \textbf{Batching}. Instead of updating $\boldsymbol{w}$ using a single data sample in each iteration, the dataset is divided into random batches of size $|B|$ and $\boldsymbol{w}$ is updated by averaging the partial derivative of the samples in a batch. %Batching improves the convergence rate as each update is averaged over all samples in the batch. It also allows the use of vectorization libraries to speed up the computation. 
With batching, the vectorized form of the update rule for the $k$th batch is given as:
$\boldsymbol{w} = \boldsymbol{w} - \alpha \frac{1}{|B|}\boldsymbol{X}^{T}_{k}(f(\boldsymbol{X}_k \times \boldsymbol{w}) - \boldsymbol{Y}_k)$. One pass through the training samples is called an epoch. 
%The dataset is usually reshuffled before each epoch.

\noindent \textbf{Learning Rate}. The learning rate $\alpha$ determines the step size of the SGD towards the minimum. A large $\alpha$ corresponds to larger steps, which may result in divergence due to overstepping the minimum. In contrast, a small $\alpha$ may increase the total iterations required. 
%One method is to dynamically tune $\alpha$ based on the accuracy of a test set. If the accuracy decreases after an epoch, we reduce the value of $\alpha$ by half and start the training again; otherwise, we run the next epoch with the same $\alpha$.   

\noindent \textbf{Termination}. When the cost does not decrease significantly for several iterations, SGD is considered to have converged to a minimum, and we can terminate. We can then use the learned coefficient vector $\boldsymbol{w}$ for inference. 

\input{logistic_regression}

\section{Reduced Precision Training}
\label{app:less_precise_training}

ML algorithms are inherently resilient to error, setting them apart from conventional algorithms that demand precise computations and high dynamic range number representations. High precision computation for training ML algorithms is unnecessary~\cite{large_scale_learning_tradeoff}.  In fact, introducing minor noise during training can enhance neural network generalization, thereby mitigating over-fitting~\cite{mlp_fault_tolerance,train_with_noise,noise_benefits_in_backprop}.

A recent trend in neural network training is to use approximate computing techniques to reduce memory requirements and improve computing efficiency~\cite{micikevicius2018mixed}. Gupta et al.~\cite{limited_precision_training_fixed_pt} show that training neural networks using 16-bit fixed-point representation with stochastic rounding can yield accuracy similar to the 32-bit floating-point format. Wang et al.~\cite{deep_training_8bit} employ a combination of 16-bit and 8-bit floating-point numbers for various deep-learning tasks. Sun et al.~\cite{HFP8_training} present an 8-bit hybrid floating-point format and demonstrate its robustness by training neural networks for various tasks while preserving accuracy.

\section{Proofs}
\label{app:proofs}

\begin{theorem}
\label{theorem:geometric_mechanism}
The geometric mechanism defined in Definition~\ref{def:geometric_mechanism} satisfies $d_{\mathcal{X}}$-privacy. 
\end{theorem}

\begin{proofsketch}
Let $x,x' \in \mathcal{X}$. Let $p_x$ and $p_{x'}$ denote the probability mass functions of $G_x$ and $G_{x'}$ respectively. Then, for some $z \in \mathcal{Y}$, we have,

% $\frac{p_x(z)}{p_{x'}(z)} =\frac{\frac{1-e^{-\epsilon}}{1+e^{-\epsilon}}\cdot e^{-\epsilon \cdot d_{\mathcal{X}}(x,z)}}{\frac{1-e^{-\epsilon}}{1+e^{-\epsilon}}\cdot e^{-\epsilon \cdot d_{\mathcal{X}}(x',z)}}$

\begin{align*}
\dfrac{p_x(z)}{p_{x'}(z)} 
&=\dfrac{\frac{1-e^{-\epsilon}}{1+e^{-\epsilon}}\cdot e^{-\epsilon \cdot d_{\mathcal{X}}(x,z)}}{\frac{1-e^{-\epsilon}}{1+e^{-\epsilon}}\cdot e^{-\epsilon \cdot d_{\mathcal{X}}(x',z)}} \\
&=\dfrac{e^{-\epsilon \cdot d_{\mathcal{X}}(x,z)}}{e^{-\epsilon \cdot d_{\mathcal{X}}(x',z)}} \\
&=e^{\epsilon \cdot (d_{\mathcal{X}}(x',z)-d_{\mathcal{X}}(x,z))}
\end{align*}

% $\frac{p_x(z)}{p_{x'}(z)} =\frac{\frac{1-e^{-\epsilon}}{1+e^{-\epsilon}}\cdot e^{-\epsilon \cdot d_{\mathcal{X}}(x,z)}}{\frac{1-e^{-\epsilon}}{1+e^{-\epsilon}}\cdot e^{-\epsilon \cdot d_{\mathcal{X}}(x',z)}}$

% $=\frac{e^{-\epsilon \cdot d_{\mathcal{X}}(x,z)}}{e^{-\epsilon \cdot d_{\mathcal{X}}(x',z)}}$
 
% $=e^{\epsilon \cdot (d_{\mathcal{X}}(x',z)-d_{\mathcal{X}}(x,z))}$ 

\noindent Using the triangle inequality for the metric $d_{\mathcal{X}}(.,.)$ and as $d_{\mathcal{X}}(x,x')=d_{\mathcal{X}}(x',x)$ , we get:

$$
\dfrac{p_x(z)}{p_{x'}(z)} \leq e^{\epsilon \cdot d_{\mathcal{X}}(x,x')}
$$

\noindent which completes the proof.

\end{proofsketch}

\section{Logistic Regression Protocol}
\label{app:lr_protocol}

We present our complete Logistic Regression protocol in Algorithm~\ref{algorithm:logistic_regression}. To implement logistic regression, we use SecureML's linear regression algorithm and add our $\Pi_{\sigmoid}$ protocol for computing the $\sigmoid$ function.

We can denote the input matrix for the $i$th layer as a $|B| \times d_i$ matrix $\boldsymbol{X_i}$ and the coefficient matrix as a $d_{i-1} \times d_i$ matrix $\boldsymbol{W_i}$. In iteration $k$, during the forward pass, the input to the first layer $\boldsymbol{X_0}$ is initialized as batch $k$ of the input data, and $\boldsymbol{X_i}$, for each layer $i$, is computed as $\boldsymbol{X_i} = f(\boldsymbol{X_{i-1}} \times \boldsymbol{W_i})$. During the back propagation, given a cost function $C(\boldsymbol{W})$, such as the cross-entropy function, we first compute $\boldsymbol{\Delta_i} = \frac{\delta C(\boldsymbol{W})}{\delta \boldsymbol{Z_i}}$ for each layer $i$, where $\boldsymbol{Z_i} = \boldsymbol{X_{i-1}} \times \boldsymbol{W_i}$. For the output layer $l$, $\boldsymbol{\Delta_l}$ is computed as: $\boldsymbol{\Delta_l} = \frac{\delta C(\boldsymbol{W})}{\delta \boldsymbol{X_l}} \odot \frac{\delta f(\boldsymbol{Z_l})}{\delta \boldsymbol{Z_l}} $; here $\frac{\delta f(\boldsymbol{Z_l})}{\delta \boldsymbol{Z_l}}$ is the derivative of the activation function and $\odot$ represents element-wise multiplication. Given $\boldsymbol{\Delta_l}$ for the output layer, $\boldsymbol{\Delta_i}$ for previous layers is computed as: $\boldsymbol{\Delta_i} = (\boldsymbol{\Delta_{i+1}} \times \boldsymbol{W_i^T}) \odot \frac{\delta f(\boldsymbol{Z_i})}{\delta \boldsymbol{Z_i}}$. The coefficients $\boldsymbol{W_i}$ in layer $i$ are then updated as: $\boldsymbol{W_i} = \boldsymbol{W_i}- \alpha \frac{1}{|B|} (\boldsymbol{X_i} \times \boldsymbol{\Delta_i})$. We use $Rec$ to denote share reconstruction.

\input{epsilon_vs_luts}

\input{comparison_with_related_works}

\section{Datasets}
\label{app:datasets}

We use the following datasets in our experiments to compare our results with the previous works~\cite{secureml, securenn} and evaluate the scalability of our protocols.

\noindent \textbf{MNIST.} The MNIST dataset~\cite{mnist} contains images of handwritten digits from `0' to `9'. The training set consists of 60,000 images, and the test set contains 10,000 images. Each sample consists of 784 features representing a $28 \times 28$ image with pixel values ranging from 0 to 255. 

\noindent \textbf{Gisette.} The Gisette dataset ~\cite{gisette, arcene_paper} is constructed from the MNIST dataset, and the task is to classify the highly confusible digits `4' and `9'. The dataset contains 13,500 samples, each with 5,000 features, with values ranging from 0 to 1,000. 

\noindent \textbf{Arcene.} The Arcene dataset~\cite{arcene, arcene_paper} contains mass-spectrometric data, and the task is distinguishing cancer from normal patterns. The dataset contains 900 instances, each with 10,000 features, with values ranging from 0 to 1000. 

\noindent \textbf{Fashion MNIST.} The dataset contains Zalando's article images in 10 classes~\cite{fashion-mnist}. The training set consists of 60,000 images, and the test set contains 10,000 images. Each sample consists of 784 features with values ranging from 0 to 255. The dataset can be used as a drop-in replacement for MNIST.

\noindent \textbf{Iris.}
The Iris dataset~\cite{iris-dataset-1} contains 3 classes of 50 instances, each where each class is a type of iris plant, and each instance consists of 4 features. 

\noindent \textbf{Adult.} The dataset consists of census data with 48,842 instances, each with 14 attributes. The classification task is to predict if an individual's income exceeds $\$50$K/year. The dataset also contains missing values. 

\section{Comparison with Related Works}
\label{app:comparison}
Table~\ref{tab:comparison_with_pika_sirnn} provides comparison of our $\textsc{Hawk}_{\text{Single}}$ and $\textsc{Hawk}_{\text{Multi}}$ protocols with Pika~\cite{waghPikaSecureComputation2022} and SIRNN~\cite{sirnn} for sigmoid function computation. Our protocols outperform both Pika and SIRNN in all settings.

%% file: logistic_regression.tex
\begin{algorithm*}[t]
%    \linespread{0.1}\selectfont
%    \small
    \caption{Logistic Regression: $\Pi_{\logistic(P_0, P_1)}$}
    %\todo{add details of what the symbols represent in the text.}}
    \label{algorithm:logistic_regression}
    \hspace{-46em} \textbf{Input:}  $\langle \boldsymbol{X} \rangle, \langle \boldsymbol{Y} \rangle $  \\
    \hspace{-48em} \textbf{Output:} $\boldsymbol{w}$ \\
    \hspace{-31.7em} \textbf{Common Randomness:} $\langle \boldsymbol{U} \rangle, \langle \boldsymbol{V} \rangle, \langle \boldsymbol{Z} \rangle, \langle \boldsymbol{V'} \rangle, \langle \boldsymbol{Z'} \rangle,$
    %$L^{\sigmoid}$
    
    \begin{enumerate}[label*=\arabic*.]

    %\hspace{-3.8em} \textbf{Input:} $\langle \boldsymbol{X} \rangle, \langle \boldsymbol{Y} \rangle $

    %\hspace{-3.8em} \textbf{Output:} $\boldsymbol{w}$

    %\hspace{-3.8em} \textbf{Common Randomness:} $\langle \boldsymbol{U} \rangle, \langle \boldsymbol{V} \rangle, \langle \boldsymbol{Z} \rangle, \langle \boldsymbol{V'} \rangle, \langle \boldsymbol{Z'} \rangle,$
    
    \item Initialize lookup table counters  $c_1, \dots, c_m = 0$, where m is the number of clients. 
    
    \item For $i \in \{0, 1\}$, $P_i$ computes $\langle \boldsymbol{E} \rangle_i = \langle \boldsymbol{X} \rangle_i - \langle \boldsymbol{U} \rangle_i$. The parties reconstruct $\boldsymbol{E}$ using $Rec(\langle \boldsymbol{E} \rangle_0, \langle \boldsymbol{E} \rangle_1)$.
    \item For $1 \leq j \leq t$
     
        \begin{enumerate}[label*=\arabic*.]
        \item Parties select a mini-batch denoted by $B_j$.
        
        \item For $i \in \{0, 1\}$, $P_i$ computes $\langle \boldsymbol{F}_j \rangle_i = \langle \boldsymbol{w} \rangle_i - \langle \boldsymbol{V}[j] \rangle$. The parties reconstruct $\boldsymbol{F}_j$ using $Rec(\langle \boldsymbol{F}_j \rangle_0, \langle \boldsymbol{F}_j \rangle_1)$. 
        \item For $i \in \{0, 1\}$, $P_i$ computes $\langle \boldsymbol{G}_{B_{j}} \rangle_i = \langle \boldsymbol{X}_{B_j} \rangle_i \times \boldsymbol{F}_j + \boldsymbol{E}_{B_j} \times \langle \boldsymbol{w} \rangle_i + \langle \boldsymbol{Z}_j \rangle_i - i \cdot \boldsymbol{E}_{B_j} \times \boldsymbol{F}_j$
%        \item For $i \in \{0, 1\}$, $P_i$ generates a random $r_i$.
        \item For $i \in \{0, 1\}$, $P_i$ calls $\Pi_{\sigmoid(P_0, P_1)}$ with inputs ($\langle \boldsymbol{G}_{B_{j}} \rangle_i$, $L^{\sigmoid}_{i}$)  and sets the output as $\langle \boldsymbol{Y}^{*}_{B_{j}} \rangle_i$.
        
%        \item If for a client k, a lookup table key is repeated: set $c_k = c_k + 1$
        
        \item  For $i \in \{0, 1\}$, $P_i$ computes $\langle \boldsymbol{D}_{B_{j}} \rangle_i = \langle \boldsymbol{Y}^{*}_{B_{j}} \rangle_i - \langle \boldsymbol{Y}_{B_{j}} \rangle_i$
        \item  For $i \in \{0, 1\}$, $P_i$ computes $\langle \boldsymbol{F}^{'}_{j} \rangle_i = \langle \boldsymbol{D}_{B_{j}} \rangle_i - \langle \boldsymbol{V}^{'}_{j} \rangle_i$. The parties reconstruct $\boldsymbol{F}^{'}_j$ using $Rec(\langle \boldsymbol{F}^{'}_j \rangle_0, \langle \boldsymbol{F}^{'}_j \rangle_1)$. 
        \item For $i \in \{0, 1\}$, $P_i$ computes $\langle \boldsymbol{\Delta} \rangle_i = \langle \boldsymbol{X}^{T}_{B_j} \rangle_i \times \boldsymbol{F}^{'}_j + \boldsymbol{E}^{T}_{B_j} \times \langle \boldsymbol{D}_{B_j} \rangle_i + \langle \boldsymbol{Z}^{'}_j \rangle_i - i \cdot \boldsymbol{E}^{T}_{B_j} \times \boldsymbol{F}^{'}_j$
        
        \item For $i \in \{0, 1\}$, $P_i$ truncates $\langle \boldsymbol{\Delta} \rangle_i$ element-wise to get $\lfloor\langle \boldsymbol{\Delta} \rangle_i\rfloor$
        
         \item For $i \in \{0, 1\}$, $P_i$ updates the coefficient vector as $\langle \boldsymbol{w} \rangle_i := \langle \boldsymbol{w} \rangle_i - \frac{\alpha}{|B|}\lfloor\langle \boldsymbol{\Delta} \rangle_i\rfloor$

        \end{enumerate}
        
    \item The parties reconstruct $\boldsymbol{w}$ using $Rec(\langle \boldsymbol{w} \rangle_0, \langle \boldsymbol{w} \rangle_1)$ and output $\boldsymbol{w}$.

    \end{enumerate}
\end{algorithm*}

%% file: epsilon_vs_luts.tex
%%%\begin{table}[h]
%%%\centering
%%%\resizebox{60mm}{!}{%
%%%\begin{tabular}{|c|c|c|c|c|}
%%%\hline
%%%      \textbf{Framework}       & \textbf{MNIST (96)} & \textbf{Iris(96)} & \textbf{Adult (85)} \\ \hline
%%%$\textsc{Hawk}_{\text{Single}}$ &   160200000    &  440550 &  65204070 \\ \hline
%%%$\textsc{Hawk}_{\text{Multi}}$ $(\epsilon_{T}=0.1)$ &  19224000     &  48060 &  6520407 \\ \hline
%%%$\textsc{Hawk}_{\text{Multi}}$ $(\epsilon_{T}=0.05)$ &   19224000    & 48060  &  10432651 \\ \hline
%%%$\textsc{Hawk}_{\text{Multi}}$ $(\epsilon_{T}=0.02)$ &   19224000    & 48060  & 10432651  \\ \hline
%%%\end{tabular}
%%%}
%%%\caption{Lookup tables required by our PPNN protocols to achieve a certain accuracy level on various datasets. $(\text{r}_{\text{Multi}} \text{= 10})$.}
%%%\label{table:luts_vs_epsilon}
%%%\end{table}

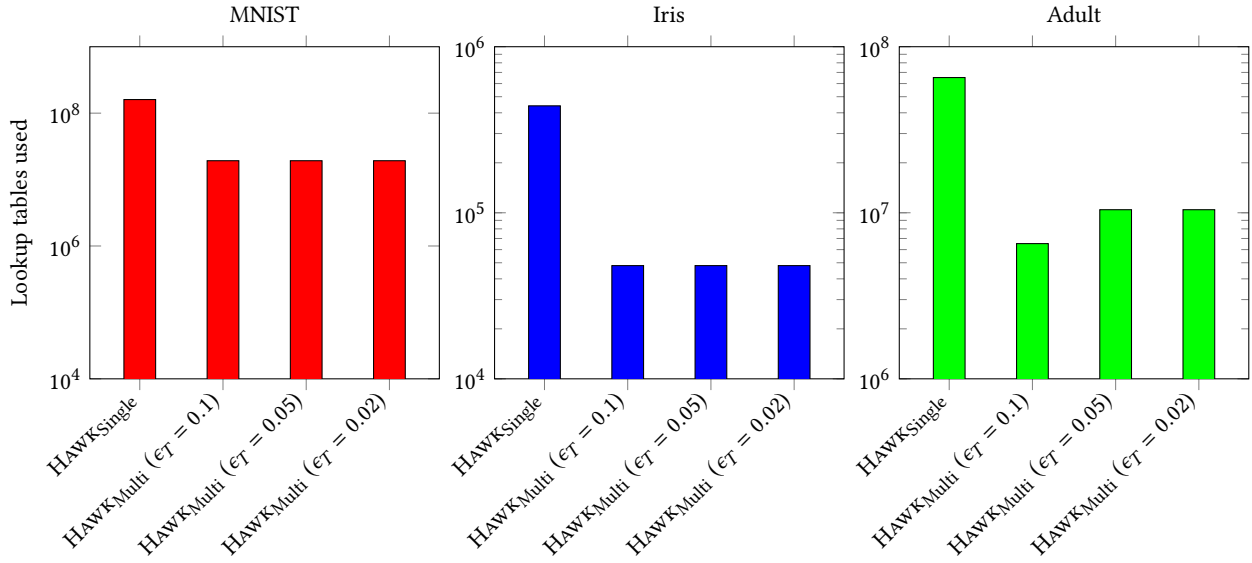
\begin{figure*}
    \centering
    % MNIST plot
    \begin{tikzpicture}
    \begin{axis}[
        width=0.35\textwidth,
        height=6cm,
        title=MNIST,
        ybar=3pt,
        bar width=12pt,
        enlarge x limits=0.2,
        symbolic x coords={$\textsc{Hawk}_{\text{Single}}$, $\textsc{Hawk}_{\text{Multi}}$ $(\epsilon_{T}=0.1)$, $\textsc{Hawk}_{\text{Multi}}$ $(\epsilon_{T}=0.05)$, $\textsc{Hawk}_{\text{Multi}}$ $(\epsilon_{T}=0.02)$},
        xtick=data,
        ylabel={Lookup tables used},
        x tick label style={rotate=45, anchor=east},
        ymin=10^4, % Adjusted to ensure the lower bound is included
        ymax=10^9, % Adjusted to ensure the upper bound is included and all ticks fit
        ymode=log,
        ytick={10^4,10^6,10^8}, % Specifying custom ytick values
    ]
    \addplot[fill=red] coordinates {($\textsc{Hawk}_{\text{Single}}$,160200000) ($\textsc{Hawk}_{\text{Multi}}$ $(\epsilon_{T}=0.1)$,19224000) ($\textsc{Hawk}_{\text{Multi}}$ $(\epsilon_{T}=0.05)$,19224000) ($\textsc{Hawk}_{\text{Multi}}$ $(\epsilon_{T}=0.02)$,19224000)};
    \end{axis}
    \end{tikzpicture}%
    % Iris plot
    \begin{tikzpicture}
    \begin{axis}[
        width=0.35\textwidth,
        height=6cm,
        title=Iris,
        ybar=3pt,
        bar width=12pt,
        enlarge x limits=0.2,
        symbolic x coords={$\textsc{Hawk}_{\text{Single}}$, $\textsc{Hawk}_{\text{Multi}}$ $(\epsilon_{T}=0.1)$, $\textsc{Hawk}_{\text{Multi}}$ $(\epsilon_{T}=0.05)$, $\textsc{Hawk}_{\text{Multi}}$ $(\epsilon_{T}=0.02)$},
        xtick=data,
        x tick label style={rotate=45, anchor=east},
        yticklabel pos=left,
        ymin=10^4, % Adjusted to ensure the lower bound is included
        ymax=10^6, % Adjusted to ensure the upper bound is included and all ticks fit
        ymode=log,
        ytick={10^4,10^5,10^6}, % Specifying custom ytick values
    ]
    \addplot[fill=blue] coordinates {($\textsc{Hawk}_{\text{Single}}$,440550) ($\textsc{Hawk}_{\text{Multi}}$ $(\epsilon_{T}=0.1)$,48060) ($\textsc{Hawk}_{\text{Multi}}$ $(\epsilon_{T}=0.05)$,48060) ($\textsc{Hawk}_{\text{Multi}}$ $(\epsilon_{T}=0.02)$,48060)};
    \end{axis}
    \end{tikzpicture}%
    % Adult plot
    \begin{tikzpicture}
    \begin{axis}[
        width=0.35\textwidth,
        height=6cm,
        title=Adult,
        ybar=3pt,
        bar width=12pt,
        enlarge x limits=0.2,
        symbolic x coords={$\textsc{Hawk}_{\text{Single}}$, $\textsc{Hawk}_{\text{Multi}}$ $(\epsilon_{T}=0.1)$, $\textsc{Hawk}_{\text{Multi}}$ $(\epsilon_{T}=0.05)$, $\textsc{Hawk}_{\text{Multi}}$ $(\epsilon_{T}=0.02)$},
        xtick=data,
        x tick label style={rotate=45, anchor=east},
        yticklabel pos=left,
        ymin=10^6, % Adjusted to ensure the lower bound is included
        ymax=10^8, % Adjusted to ensure the upper bound is included and all ticks fit
        ymode=log,
        ytick={10^6,10^7,10^8}, % Specifying custom ytick values
    ]
    \addplot[fill=green] coordinates {($\textsc{Hawk}_{\text{Single}}$,65204070) ($\textsc{Hawk}_{\text{Multi}}$ $(\epsilon_{T}=0.1)$,6520407) ($\textsc{Hawk}_{\text{Multi}}$ $(\epsilon_{T}=0.05)$,10432651) ($\textsc{Hawk}_{\text{Multi}}$ $(\epsilon_{T}=0.02)$,10432651)};
    \end{axis}
    \end{tikzpicture} 
    \caption{Lookup tables required by our PPNN protocols to achieve a certain accuracy level i.e. 96\% for MNIST, 85\% for Adult, and 96\% for Iris dataset.$(\text{r}_{\text{Multi}} \text{= 10})$.}
    \label{fig:luts_vs_epsilon}
\end{figure*}

%% file: comparison_with_related_works.tex
\begin{table*}[ht]
\centering
\begin{tabular}{|l|c|c|c|c|c|c|c|}
\hline
             & \textbf{Mode}        & $\mathbf{10^0}$ & $\mathbf{10^1}$ & $\mathbf{10^2}$ & $\mathbf{10^3}$ & $\mathbf{10^4}$ & $\mathbf{10^5}$ \\ \hline
Pika         & \multirow{4}{*}{LAN} & 0.0007                         & 0.001                          & 0.006                          & 0.059                          & 0.539                          & 5.329                          \\ \cline{1-1} \cline{3-8} 
SIRNN        &                      & \_                             & 0.115                          & 0.116                          & 0.147                          & 0.289                          & 2.089                          \\ \cline{1-1} \cline{3-8} 
$\textsc{Hawk}_{\text{Single}}$ &                      & 0.0002                         & 0.0004                         & 0.0004                         & 0.0013                         & 0.0060                         & 0.0482                         \\ \cline{1-1} \cline{3-8} 
$\textsc{Hawk}_{\text{Multi}}$  &                      & 0.0008                         & 0.0016                         & 0.0038                         & 0.0209                         & 0.1855                         & 1.8193                         \\ \hline
Pika         & \multirow{4}{*}{WAN} & 0.183                          & 0.195                          & 0.198                          & 0.356                          & 1.375                          & 9.030                          \\ \cline{1-1} \cline{3-8} 
SIRNN        &                      & \_                             & 7.683                          & 7.690                          & 7.816                          & 9.313                          & 32.702                         \\ \cline{1-1} \cline{3-8} 
$\textsc{Hawk}_{\text{Single}}$ &                      & 0.050                          & 0.059                          & 0.059                          & 0.062                          & 0.066                          & 0.116                          \\ \cline{1-1} \cline{3-8} 
$\textsc{Hawk}_{\text{Multi}}$  &                      & 0.179                          & 0.180                          & 0.185                          & 0.192                          & 0.318                          & 2.160                          \\ \hline
\end{tabular}
\caption{Comparison of $\textsc{Hawk}$ with Pika and SIRNN semi-honest protocols. Time is in seconds and the computation is a sigmoid function on a batch of $\mathbf{10^0}$, $\mathbf{10^1}$, $\mathbf{10^2}$, $\mathbf{10^3}$, $\mathbf{10^4}$, and $\mathbf{10^5}$ in both the LAN and WAN settings. }
\label{tab:comparison_with_pika_sirnn}
\end{table*}